\documentclass[
preprint,
amssymb,
 aps,
 pre,
 floatfix,
 superscriptaddress,
 footinbib,
 sort&compress, numbers, merge
]{revtex4-1}

\usepackage[fleqn]{amsmath}
\usepackage{amssymb}
\usepackage{array}
\usepackage{natbib}
\usepackage{amssymb}
\usepackage[figuresright]{rotating}

\usepackage[usernames,dvipsnames]{xcolor}

\begin{document}




\title{From wrinkling to global buckling
    of a ring on a curved substrate}


\author{R. Lagrange}
\affiliation{Department of Mathematics, Massachusetts Institute of Technology,
Cambridge, MA 02139, USA}
\author{F. L\'opez Jim\'enez}
\affiliation{Department of Civil and
Environmental Engineering, Massachusetts Institute of Technology,
Cambridge, MA 02139, USA}
\author{D. Terwagne}
\thanks{Current address: Department of Physics, Facult\'e des
Sciences, Universit\'e Libre de Bruxelles (ULB), Bruxelles 1050,
Belgium.}
\affiliation{Department of Civil and
Environmental Engineering, Massachusetts Institute of Technology,
Cambridge, MA 02139, USA}
\author{M. Brojan}
\thanks{Current address: Faculty of Mechanical Engineering,
University of Ljubljana, Slovenia.}
\affiliation{Department of Civil and
Environmental Engineering, Massachusetts Institute of Technology,
Cambridge, MA 02139, USA}
\author{P. M. Reis}
\thanks{Email: preis@mit.edu}
\affiliation{Department of Civil and
Environmental Engineering, Massachusetts Institute of Technology,
Cambridge, MA 02139, USA}
\affiliation{Department of Mechanical Engineering, Massachusetts Institute of Technology,
Cambridge, MA 02139, USA}


\begin{abstract}

We present a combined analytical approach and numerical study on the
stability of a ring bound to an annular elastic substrate, which
contains a circular cavity.  {\color{black}{The system is loaded by depressurizing the inner cavity.}}
The ring is modeled as an Euler-Bernoulli
beam and its equilibrium equations are derived from the mechanical
energy which takes into account both stretching and bending
contributions. The curvature of the substrate is considered
explicitly to model the work done by its reaction force on the ring.
We distinguish two different instabilities: periodic wrinkling of
the ring or global buckling of the structure. Our model provides an
expression for the critical pressure, as well as a phase diagram
that rationalizes the transition between instability modes. Towards
assessing the role of curvature, we compare our results for the
critical stress and the wrinkling wavelength to their planar
counterparts. We show that the critical stress is insensitive to the
curvature of the substrate, while the wavelength is only affected due to
the permissible discrete values of the azimuthal wavenumber imposed by the
geometry of the problem. Throughout, we contrast our analytical
predictions against finite element simulations.

\end{abstract}





\maketitle

\section{Introduction}
\label{sec:introduction}

Wrinkling is a stress-driven mechanical instability that occurs when
a stiff and slender surface layer, bonded to a compliant substrate, is
subject to compression. This universal instability phenomenon is found
in numerous natural and technological/engineering examples, over a wide range of length scales, including: carbon
nanotubes~\citep{Lourie98}, pre-stretched elastomers used in
flexible electronics applications~\citep{Kim11}, human
skin~\citep{Yin10}, drying fruit~\citep{Yin09}, surface morphology
of the brain~\citep{Budday14} and mountain topographies generated
due to tectonic stresses~\citep{Price90,Huddleston93}.

Over the past decade, there has been an upsurge of interest in the
study of the mechanics of wrinkling, along with a change of paradigm
in regarding surface instabilities as an opportunity for
functionality, instead of a first step in the route to structural
failure~\citep{Genzer06, Li12}. The first mechanical studies of
wrinkling were motivated by the stability of sandwich panels
\citep{Allen69}, used in lightweight structural applications, in
which the core acts as a soft substrate for the much stiffer skin.
More recently, \cite{Bowden98} showed how the wrinkling of a thin
film on an elastomeric substrate can be used to produce complex
self-organized patterns. Their seminal work has instigated the
realization of wrinkling through several different actuation
mechanisms, including thermal mismatch \citep{Huck00}, tissue
growth/atrophy \citep{BenAmar05,Li11c,Budday14}, swelling by a
liquid \citep{Chan06Soft} or vapor solvent \citep{Breid09}, and
pneumatics \citep{Terwagne14}. The opportunities in applications opened by such a
wide range of external stimuli have enabled the usage of wrinkling
in photonics~\citep{Kim12},
optics~\citep{Chan06Advanced}, self-assembly~\citep{Yoo02},
microfluidics~\citep{Yin12} and morphogenesis \citep{Efimenko05}.

In order to provide a theoretical background to these recent
developments, several authors have built on the pioneering work of
\cite{Allen69}, who first provided close form solutions for the
critical stress and wavelength obtained when an initially straight
beam, adhered to an infinite plane substrate, is placed under a state
of uniaxial compression. \cite{Chen04} extended this work to
consider the case of a plate adhered to a flat substrate under
equi-biaxial compression and performed a nonlinear analysis of the
F{\"o}ppl–-von K\'arm\'an equations \citep{Landau59,Timoshenko61}.
\cite{Huang05} further refined these efforts by considering the
effect of a finite substrate. Both  studies showed the existence of
multiple buckling modes associated with the same value of critical
stress. The stability of these modes under different loadings
conditions has been addressed by
\cite{Audoly08a,Audoly08b,Audoly08c}, who produced a stability
diagram covering the evolution from low to high values of
overstress. However, experiments by \cite{Cai11} found disagreement
at low values of overstress, suggesting that a finite intrinsic
curvature of their experimental system, even if small, may play an important role
in dictating  pattern selection.

Early studies of wrinkling on curved substrates, as in the flat
configuration, were also motivated by a structural problem; in this
case, in the context of the stability of the outer shell of rockets
\citep{Kachman59,Seide61,Seide62}. More recent studies that consider
instabilities as a possible source of functionality have led to
applications of curved configurations  in adhesion \citep{Kundu11},
microfluidics \citep{Mei2010}, morphogenesis of microparticles
\citep{Yin14}, optics \citep{Breid13} and aerodynamic drag reduction
\citep{Terwagne14}. Curvature also plays a relevant role in the
growth of biological systems \citep{Li11a}. Despite these important
emerging applications, the mechanics of wrinkling on curved
substrates remains poorly understood, when compared to the planar
counterpart.

Systematic Finite Element simulations of wrinkling in curved systems have been
performed \citep{Yin09,Yin10,Li11b,Cao12} that highlighted a complex
pattern formation process. These numerical studies also suggested the possibility for curvature
to affect the selected patterns and modify the relevant
characteristic length scales, which calls for a robust theoretical
backing. Analytical predictions are challenged by the difficulty of
modeling the stiffness of the substrate, even in two-dimensional
configurations. \cite{Cheng96} and \cite{Cai11} used the stiffness
provided by \cite{Allen69} for the flat case, such that their model
therefore neglects the contribution of curvature on the response of
the substrate. \cite{Yin09} used the prediction provided by
\cite{Brush75}, which accounts for curvature but does not consider
its influence on the wrinkling wavelength and their prediction does
not converge to the classical planar case when the curvature tends to
zero. As such, there is a need to quantify the effect of curvature
on the stiffness of the substrate and its subsequent influence on
wrinkling.

Here, to the best of our knowledge, we provide the first analytical
work that accounts for both the curvature of a (2D) shell-substrate
system, as well as the finite size of the substrate. As an initial
step, we focus our study on a curved film adhered to a cylindrical substrate,
instead of dealing with non-zero Gaussian curvature geometries, which is left for a future study. We
assume axial--symmetry to further simplify the system to
the 2D problem of a ring on an annular substrate.  {\color{black}{Mechanical loading is applied by 
depressurizing a circular cavity inside the substrate, which places the system under a state of compression.}} This
geometry is motivated by recent experiments
\citep{Terwagne14} that demonstrated the usage of wrinkling on
spherical samples for switchable and tunable aerodynamic drag
reduction. In our simplified 2D system, we solve the elasticity
problem for the substrate and derive a close form expression for its
stiffness, which is then used in the stability analysis of the ring to quantify the buckling patterns.

The paper is organized as follows: In \S~\ref{Subsection
Presentation}, we introduce our system along with its material and
geometrical parameters. We also describe the possible instability
modes, and present a simplified phase diagram, with the aim of
providing physical insight on the problem. In
\S~\ref{sec:analytical}, we then introduce the kinematics of the ring
attached to the substrate and determine the stiffness of the substrate.
We proceed by defining a strain energy that includes both bending
and stretching of the ring, as well as the effect of the substrate.
Energy minimization yields the equilibrium equations of the problem.
An asymptotic expansion is then used to calculate the principal
solution and the bifurcation at the onset of instability. In
\S~\ref{sec:fem}, we describe the finite element simulations that we
have performed for this same system.

The results of our investigation are presented in
\S~\ref{sec:results}. Throughout, we directly compare the analytical
predictions to the numerical simulations. We start with the
fundamental solution and the critical conditions that lead to
instability. We then construct a phase diagram which rationalizes
the dependence of the instability modes on the governing parameters.
The results for our system are then quantitatively compared to those
for wrinkling of a film on a planar substrate, highlighting the
effect of curvature. Finally, \S~\ref{sec:discussion} summarizes our
findings and provides perspectives for potential extensions of our
work in future studies.

\section{Definition of the problem}\label{Subsection Presentation}

\begin{figure}[h!]
    \begin{center}
\includegraphics[width=0.75\columnwidth]{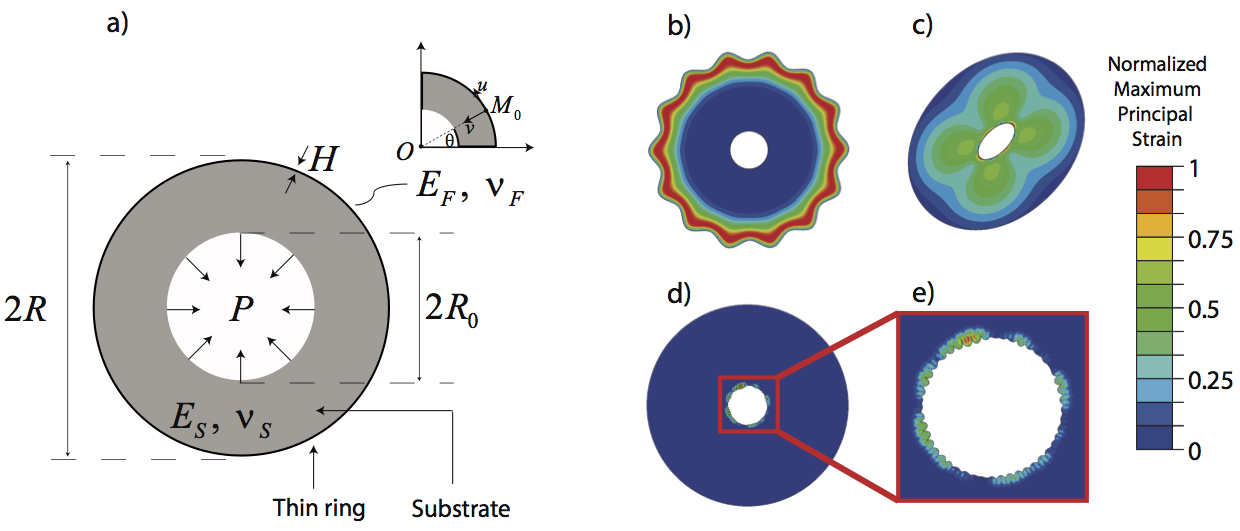}
\end{center}
\caption{(a) Schematic diagram of our system: a ring is bound to a
curved substrate which contains a circular cavity. The system is
loaded by applying a pressure differential between the inside of the
cavity and the outside of the ring. (b-d) Representative
examples of the three possible instability modes of a ring on a curved
substrate which contains a cavity that is depressurized. (b)
Wrinkling mode ($h = 10^{-2}$ and $\xi = 10^3$), (c) global buckling mode
 ($h = 10^{-2}$ and $\xi = 10^6$) and (d) Biot mode ($h = 10^{-2}$ and $\xi = 10^2$). (e) is a
zoom in of (d) that exhibits the deformation of the surface of the
inner cavity in the Biot mode. {\color{black}{ The adjacent colorbar applies to pictures b-e) and refers to the
maximum principal component of the strain tensor of the mode, which has been
normalized by the maximum value of each configuration.}}  } \label{Fig0}
\end{figure}

We study the stability of a thin elastic ring, bound to an equally
curved 2D substrate that contains an inner cavity, a schematic
diagram of which is presented in Fig. \ref{Fig0}(a). {\color{black}{The system is initially at equilibrium, with identical pressures inside and outside of the
sample. Motivated by recent experiments on spherical specimens \citep{Terwagne14},
the system is then loaded by applying a depressurization, P, to the inner cavity.}} The thickness of the ring
is $H$, its Young's modulus $E_F$ and its Poisson's ratio $\nu_F$.
We refer to $\overline{E_F}=E_F/(1-{\nu_F}^2)$ as the reduced
Young's modulus of the filmß. The substrate is made of a linearly
elastic material with Young's modulus $E_S$, Poisson's ratio $\nu_S$
and reduced Young's modulus $\overline{E_S}=E_S/(1-{\nu_S}^2)$. The
thickness of the substrate is $R-R_0$, where $R_0$ is the radius of
the inner cavity.

For convenience, we now introduce new rescaled quantities
 to reduce the number of parameters of the
problem. As such, we use $R$ and $\overline{E_F}$ to normalize
lengths and pressures and define
\begin{equation}
h = \frac{H}{R},\,\,\,\beta  = \frac{{{R_0}}}{R},\,\,\,\xi  =
\frac{{{\overline{E_F}}}}{{\overline {{E_S}} }},\,\,\,p =
\frac{P}{{\overline {{E_F}} }},
\end{equation}
as the dimensionless thickness, cavity size, stiffness ratio, and
pressure, respectively.

 {\color{black}{The principal solution corresponds to an axisymmetric deformation that leads to a decrease of both the inner and outer radii. As the depressurization increases, the onset of instability is reached. In Fig.~\ref{Fig0}(b-e) we show representative results obtained
from Finite Element Modeling (FEM), of the three possible
instability configurations of the ring-substrate system, for
different values of the dimensionless ring thickness, $h$, and ratio
of stiffness, $\xi$. For these results, all the other mechanical properties were kept constant: the Poisson's ratios of the film and substrate are $\nu_F=\nu_S=0.5$, and the dimensionless size of the cavity is $\beta = 0.2$. In the simulations, this is achieved by fixing $R = 100$ units of length and $E_S = 1$ units of
pressure, while changing the values of $H$, $R_0$ and $E_F$
accordingly. More details of our numerical simulations are provided
in \S \ref{sec:fem}.}}

 {\color{black}{The first mode, a representative example of which is shown in Figure~\ref{Fig0}(b) for $h = 10^{-2}$ and $\xi = 10^3$, corresponds to periodic wrinkling of the film with a well defined wavelength. The displacements are localized in the region close to the film. The second instability mode,  for example at $h = 10^{-2}$ and $\xi = 10^6$ in Figure~\ref{Fig0}(c), corresponds to a global buckling of the
structure, where both the ring and the cavity
deform into an ellipse such that the wavelength is  $\lambda=\pi R$.}}

In addition to these two instability modes (wrinkling and global
buckling), we have also numerically observed an instability on the
inner surface of the cavity. However, this third mode does not
affect the ring and is only found for low values of cavity size and
stiffness ratio ($\xi = 10^2$ and $h = 10^{-2}$ in
Fig.~\ref{Fig0}c-d). This instability was first discussed by
\cite{Biot65} and we therefore refer to it as the \emph{Biot mode};
it is local in nature and only depends on the compressive strain at
the inner surface. This type of instability mode has been recently studied
in the case of elastomeric materials with voids by \cite{Michel07}
and \cite{Cai10}. Understanding the specifics of this Biot mode is
however outside the scope of our work and we shall not take it into
account in our analytical model and systematic numerical
investigation.

\begin{figure}[h!]
    \begin{center}
\includegraphics[width=0.75\columnwidth]{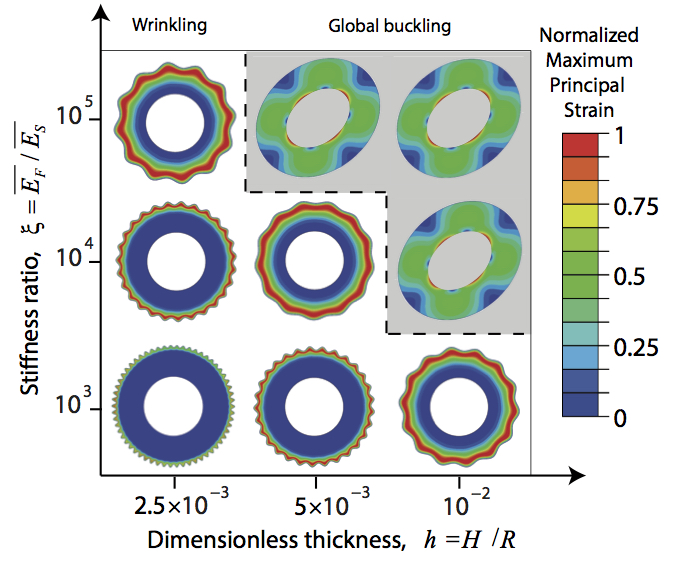}
\end{center}
\caption{Schematic phase diagram of the instability modes of the
system, obtained for a critical value of the pressure differential:
i) wrinkling of the ring or ii) global buckling of the structure
(shaded regions). The primary parameters that govern this transition
are the stiffness and thickness ratios:
$\xi=\overline{E_F}/\overline{E_S}$ and $h=H/R$, respectively. The
depicted examples from FEM simulations are for an incompressible
film and substrate, $\nu_F=\nu_S=0.5$. They were obtained for
$R=100$ units of length and $E_S=1$ units of pressure, while varying
$H$ and $E_F$. {\color{black}{ The colorbar refers to the
maximum principal component of the strain tensor of the mode, which has been
normalized by the maximum value of each configuration.}}  }\label{Sketch_Phase_Diag}
\end{figure}

A schematic phase diagram of our system is provided in Fig.~\ref{Sketch_Phase_Diag}.
For low values of $h$ and $\xi$, the ring wrinkles with a short wavelength.
As either $h$ or $\xi$ are increased, the wavelength also increases.
Once these parameters reach a critical value, represented by the dashed line
in Fig.~\ref{Sketch_Phase_Diag}, the instability transitions from wrinkling to global buckling.
In what follows, we focus on rationalizing how the wavelength of
the wrinkling mode, and the threshold value for the transition to global
buckling, evolve with the elastic and geometrical parameters of the system.

\section{Analytical model}
\label{sec:analytical}

The ring is treated as an Euler-Bernoulli beam. The effect of the substrate is
modeled as a restoring force that acts on the ring and is determined
by solving the elasticity problem of the substrate with adequate
boundary conditions. Minimization of the potential energy provides
the equilibrium equations of the problem, which are solved using an
asymptotic expansion that yields the principal and bifurcated
solutions.

\subsection{Kinematics, energy formulation and equations of equilibrium}

We model the ring as an extensible Euler-Bernoulli beam made of an
homogeneous and isotropic material. Polar coordinates are used to
track the position of the ring center-line, $C$.  The initial configuration
of the ring, prior to depressurization, is assumed to be circular.
The origin, $O$, is located at the center of the cavity, and the initial and equilibrium configurations of an arbitrary point of $C$ are represented by $M_0$ and $M$, respectively.
Vectors are
expressed in the physical base $\left(
{{{\bf{e}}_{\bf{r}}},{{\bf{e}}_{\bf{\theta }}}} \right)$, derived
from the polar coordinates $(r,\theta)$.
The initial position of $C$ is
${\bf{O}}{{\bf{M}}_0} = \left( {R,0} \right)$, as shown in the inset
of Fig.\ref{Fig0}(a). When the system is loaded by depressurizing
the cavity, $C$ deforms into a new configuration given by the position
vector ${\bf{OM}} = R\left( {1 + v\left( \theta \right),u\left(
\theta \right)} \right)$, where $v$ and $u$ are the dimensionless
radial and orthoradial displacements, respectively.

The infinitesimal arclength of $C$ in the initial and deformed
configurations are denoted by $d{s_0} = \left|
{d{\bf{O}}{{\bf{M}}_0}} \right|$ and $ds = \left| {d{\bf{OM}}}
\right|$, respectively. Moreover, defining the tangent vector
${\bf{T}} = {{d{\bf{OM}}} \mathord{\left/
 {\vphantom {{d{\bf{OM}}} {ds}}} \right.
 \kern-\nulldelimiterspace} {ds}}$, allows us to express the curvature of $C$ in the deformed configuration as ${\kappa  \mathord{\left/
 {\vphantom {\kappa  R}} \right.
 \kern-\nulldelimiterspace} R} = \left| {{{d{\bf{T}}} \mathord{\left/
 {\vphantom {{d{\bf{T}}} {ds}}} \right.
 \kern-\nulldelimiterspace} {ds}}} \right|$.
Here, $\kappa$ is dimensionless and can be written in terms of $v$ and
$u$ as
\begin{equation}\label{eqCurvature}
\kappa  = 1 + {\mkern 1mu} \left( { - 1 + 2u' + 2{\mkern 1mu} v}
\right)v'' + {v^2} - {\mkern 1mu} v - {\mkern 1mu} {u'^2} + \left( {
- {\mkern 1mu} u + {\mkern 1mu} v'} \right)u'' + \frac{1}{2}\left(
{{{v'}^2} - {u^2}} \right) + {\rm{h}}{\rm{.o}}{\rm{.t}}{\rm{.}},
\end{equation}
where the prime notation represents derivation with respect to
$\theta$ and  high order terms (h.o.t.) are neglected
under the assumption of small displacements and moderate rotations. We now define the elongation of the ring as $e = {{ds} \mathord{\left/
 {\vphantom {{ds} {d{s_0}}}} \right.
 \kern-\nulldelimiterspace} {d{s_0}}}$ to express the stretching
deformation, $\eta  = {{\left( {{e^2} - 1} \right)} \mathord{\left/
 {\vphantom {{\left( {{e^2} - 1} \right)} 2}} \right.
 \kern-\nulldelimiterspace} 2}$, in terms of $v$ and $u$ as
\begin{equation}\label{eqStretching}
\eta  = \frac{1}{2}{\mkern 1mu} \left[ {{{u'}^2} + {\mkern 1mu}
{\mkern 1mu} {v^2} + {\mkern 1mu} {\mkern 1mu} {{\left( {u - v'}
\right)}^2}} \right] + {\mkern 1mu} \left( {1 + {\mkern 1mu} v}
\right)u' + v,
\end{equation}
so that the hoop stress in the film is
$\sigma_0=\overline{E_F}\eta$.

Following Euler-Bernoulli beam theory \citep{Timoshenko61}, the
total energy of deformation $\mathcal{E}$ of the ring is the sum of
a stretching energy $\mathcal{E}_{S}$ and a bending energy
$\mathcal{E}_{B}$,
\begin{equation}
\mathcal{E} ={\mathcal{E}_{S}} +
{\mathcal{E}_{B}}=\int\limits_0^{2\pi R} {\overline{\mathcal{E}}
d{s_0}},
\end{equation}
with
\begin{subequations}
\begin{align}
{\mathcal{E}_{S}} &= \int\limits_0^{2\pi R} {\frac{{\overline
{{E_F}}
H}}{2}{\mkern 1mu} {\mkern 1mu} {\eta ^2}} d{s_0},\\
{\mathcal{E}_{B}} &= \int\limits_0^{2\pi R} {\frac{\overline{E_F}
H^3}{{24{R^2}}}{{\left( {\kappa - 1} \right)}^2}} d{s_0},
\end{align}
\end{subequations}
and $\overline{\mathcal{E}}$ is the energy of deformation per unit
length of the initial configuration of the ring. Assuming that the
reaction force of the substrate derives from a potential
$\int\limits_0^{2\pi R} {\overline{W} d{s_0}}$, the equilibrium
states of the ring are the solutions of
\begin{equation}\label{EnergyVariation}
{\delta \mathcal{E}_{S}} + {\delta \mathcal{E}_{B}} -
\int\limits_0^{2\pi R} {\delta\overline{W} d{s_0}}=0,
\end{equation}
where $\delta \mathcal{A}$ is the variation of quantity
$\mathcal{A}$, for an arbitrary displacement field $R\left(\delta v,
\delta u\right)$, which is $2\pi$ periodic. The computation of the
variations in Eq.~(\ref{EnergyVariation}) leads to the Euler-Lagrange equations for the equilibrium of the ring,
{\color{black}{
\begin{subequations}\label{eqEuler}
\begin{align}
\frac{{\partial {\overline{\mathcal{E}}}}}{{\partial v}} - {\left(
{\frac{{\partial {\overline{\mathcal{E}}}}}{{\partial v^{'}}}}
\right)^{'}} + {\left( {\frac{{\partial
{\overline{\mathcal{E}}}}}{{\partial v^{''}}}} \right)^{''}}-
\frac{{\partial {\delta \overline W}}}{{\partial \delta v}} &= 0,\label{eqEuler1}\\
\frac{{\partial {\overline{\mathcal{E}}}}}{{\partial u}} - {\left(
{\frac{{\partial {\overline{\mathcal{E}}}}}{{\partial u^{'}}}}
\right)^{'}} + {\left( {\frac{{\partial
{\overline{\mathcal{E}}}}}{{\partial u^{''}}}} \right)^{''}} -
\frac{{\partial {\delta \overline W}}}{{\partial \delta u}} &=
0, \label{eqEuler2}
\end{align}
\end{subequations}}}
along with static boundary conditions that are naturally satisfied
due to the $2 \pi$ periodicity condition on the displacements $v$
and $u$. All derivative terms in Eq.~(\ref{eqEuler2}) are explicitly reported in Appendix~\ref{Appendix Euler Lagrange}.

\subsection{Asymptotic expansion and reactive force of the substrate}
\label{Subsection Substrate}

We seek  a solution of Eq.~(\ref{eqEuler}) as an expansion of the
form
\begin{subequations}\label{eqKoiter}
\begin{align}
v &= {v_0} + \varepsilon A\sin \left( {m\theta } \right) + O\left(
{{\varepsilon ^2}} \right),\\
u &= \varepsilon B\cos \left( {m\theta } \right) + O\left(
{{\varepsilon ^2}} \right),
\end{align}
\end{subequations}
where $(v_0,0)$ corresponds to a radial pre-buckling deformation and
$\varepsilon(A\sin \left( {m\theta } \right), B\cos \left( {m\theta
} \right))$ represents an instability of azimuthal wavenumber $m$.
From the requirement of $2\pi$ periodic functions $v$ and $u$, $m$
has to be an integer. Also, we consider $m>1$ since $m=1$
corresponds to a solid body translation of the ring. The instability
displacement field has amplitudes $\varepsilon A$ and $\varepsilon
B$, where $\varepsilon$ is a small parameter.

Before substituting Eq.~(\ref{eqKoiter}) into Eq.~(\ref{eqEuler})
and solving at each order in $\varepsilon$, we first need to
determine the work, $\delta \overline{W}$, done by the reaction
force, ${\bf{F}}=-  \left( {{\sigma}{{\bf{e}}_ \bot } +
{\tau}{{\bf{e}}_\parallel }} \right)$, that the substrate exerts on
the ring. The normal stress $ {\sigma}$ and the tangential stress $
{\tau}$ at the interface between the ring and the substrate are
computed by solving the corresponding two-dimensional elasticity
problem using an Airy function, with pressure $P$ at $r=R_0$ and the
displacement field given by Eq.~(\ref{eqKoiter}), at $r=R$.

{\color{black}{ It is worth to note that in our computation, we enforce the continuity of the displacement
and stress fields at the interface ring/substrate, similarly to \cite{Mei2011}. This is
a main difference with previous studies of wrinkling surfaces, e.g. in \cite{Seide62}, who assumed zero shear stress and continuity of the normal
stress. Such an approach yields a stretching
energy in the ring much larger than the bending energy, in
contradiction to what is expected in the wrinkling of a thin film  \citep{Audoly08a}.
Our explicit solution for the boundary value problem of the substrate is
reported in \ref{Appendix Response of the substrate}. }}

In short, we
find that the stresses ${\sigma}$ and $\tau$ at the interface are
\begin{subequations}\label{eqDimlessStress}
\begin{align}
\frac{ \sigma}{\overline{E_F}} &={k_0}{v_0} + k\varepsilon A\sin
(m\theta ) -
\gamma p,\label{sigma_expression}\\
\frac{ \tau }{\overline{E_F}} &= \mu \varepsilon B\sin (m\theta),
\end{align}
\end{subequations}
with the following governing parameters
\begin{subequations}\label{eqDimlessStiff}
\begin{align}
{k_0} &= \frac{{\left( {1 - {\nu _S}} \right)\left( {1 - {\beta ^2}}
\right)}}{{1 - 2{\mkern 1mu} {\nu _S} + {\beta ^2}}}\frac{1}{\xi
},\\
k &= \frac{{\left( {1 - {\nu _S}} \right)}}{2}\left( {{S_A}{\mkern
1mu}  + {S_B}\frac{B}{A}} \right)\frac{1}{\xi },\label{Eq_k}\\
\mu &= {\mkern 1mu} {\mkern 1mu} \frac{{\left( {1 - {\nu _S}}
\right)}}{2}\left( {{S_B}{\mkern 1mu} \frac{A}{B} + {T_B}{\mkern
1mu} } \right)\frac{1}{\xi },\\
\gamma  &= \frac{{2{\beta ^2}\left( {1 - {\nu _S}} \right)}}{{1 -
2{\mkern 1mu} {\nu _S} + {\beta ^2}}}. \label{eqGAmma}
\end{align}
\end{subequations}
Here, $k_0$, $k$, and $\mu$ are the dimensionless pre-wrinkling,
wrinkling and shear stiffnesses of the substrate, respectively. The
coefficient $\gamma$ quantifies the effect of pressure, $p$, on the
ring through the term $-\gamma p$ in Eq.~(\ref{sigma_expression}).
The value of $\gamma$ is smaller than one, reflecting the fact that
the transmitted pressure decreases within the substrate. There are
however two exceptions for which $\gamma=1$: the limit where there
is no substrate, $\beta \to 1$, and the limit of a perfectly
incompressible substrate, $\nu_S \to 0.5$, in which the volumetric
pressure remains constant throughout the substrate. The quantities
$S_A$, $S_B$ and $T_B$ used in Eq. (\ref{eqDimlessStiff}) are
functions of both the cavity size $\beta$ and the azimuthal
wavenumber $m$, and their full expressions are reported in
\ref{Appendix Response of the substrate}.

{\color{black}{From the solution of the linear elasticity problem for the substrate (see \ref{Appendix Response of the substrate}), the reaction force, $\bf{F}$, has a
constant direction along ${\bf{e}}_r$. However, based on physical
intuition, one would expect the restoring force to change its direction as the ring deforms.
Accounting for this scenario in full would have required solving the elasticity problem
for the substrate, with nonlinear kinematics. Here, for simplicity, we assume that the
magnitude of the restoring force is given by the solution of the linear elasticity problem,
while the direction of the force is given by the vectors ${\bf{e}}_\bot$ and ${\bf{e}}_\parallel$. }} Depending on which configuration of the ring is used (initial or
deformed), two cases need to be considered to define the normal and
tangential vectors ${\bf{e}}_\bot$ and ${\bf{e}}_\parallel$,
respectively. If we define these vectors from the initial
configuration, then $\left( {{{\bf{e}}_ \bot },{{\bf{e}}_\parallel
}} \right) = \left( {{{\bf{e}}_{\bf{r}}},{{\bf{e}}_{\bf{\theta }}}}
\right)$, so that the reaction force ${\bf{F}}$ keeps a constant
direction while the ring deforms. If we use the deformed
configuration, then ${\bf{F}}$ is modeled as a follower force whose
direction changes with the ring deformation. In this case, $\left(
{{{\bf{e}}_ \bot },{{\bf{e}}_\parallel }} \right) = \left(
{{\bf{N}},{\bf{T}}} \right)$, where ${\bf{N}}$ is the inward normal
vector orthogonal to the ring center-line, in its deformed state. In
order to take both of these options into account in the same model,
we define a parameter, $\chi$, which can take the values  $\chi=1$
or $\chi=0$, when either the undeformed or deformed configurations
are used. Thus, we express the reaction force and its elementary
work per unit length of the ring as
\begin{subequations}\label{eqForceOnFilm2models}
\begin{align}
{\bf{F}} &=  -  \left[ {\left( {{\sigma }{{\bf{e}}_{\bf{r}}} + {\tau
}{{\bf{e}}_{\bf{\theta }}}} \right)\chi + \left( {{\sigma }{\bf{N}}
+ {\tau }{\bf{T}}} \right)\left( {\chi - 1} \right)} \right],\\
\delta  \overline{W} &=  {{\bf{F}} \cdot R\left( {\delta v
{{\bf{e}}_{\bf{r}}} + \delta u {{\bf{e}}_{\bf{\theta}}}} \right)}.
\end{align}
\end{subequations}
where $\cdot$ is the Euclidean dot product.

\subsubsection{The principal solution: order 0 in $\varepsilon$}

We now proceed to obtain the principal solution and the critical
instability modes by substituing Eq.~(\ref{eqKoiter}) into
Eq.~(\ref{eqEuler}) and solving the resulting equations for each
order of $\varepsilon$. At order 0,  Eq.~(\ref{eqStretching}) for
the stretching deformation $\eta$ yields
$\eta_0=v_0=\sigma_0/\overline {E_F}$ and the linear approximation
of the Euler-Lagrange Eq.~(\ref{eqEuler1})
\begin{equation}\label{eqV0}
{v_0} = \frac{{\gamma p}}{{h + {{{h^3}} \mathord{\left/
 {\vphantom {{{h^3}} {12}}} \right.
 \kern-\nulldelimiterspace} {12}} + {\mkern 1mu} {k_0}}}= \frac{{{\sigma _0}}}{{\overline
 {E_F}}},
\end{equation}
relates the dimensionless pressure, $p$, and the dimensionless hoop
stress, $\sigma_0/\overline {E_F}$, in the ring. Note that
Eq.~(\ref{eqEuler2}) is automatically satisfied at order
0 in $\varepsilon$. 
In the absence of a substrate, the ratio $\sigma_0/P$ reads
\begin{equation}\label{LimitNosubstratSigma}
\mathop {\lim }\limits_{\beta  \to 1} \left( {\frac{{{\sigma
_0}}}{P}} \right) = \frac{1}{h} + O\left( h \right),
\end{equation}
and in the limit of the ring that is much stiffer than the substrate we obtain
\begin{equation}\label{LimitXiInfiniSigma}
\mathop {\lim }\limits_{\xi  \to \infty } \left( {\frac{{{\sigma
_0}}}{P}} \right) = \frac{\gamma }{h} + O\left( h \right),
\end{equation}
so that ${\frac{{{\sigma _0}}}{P}}<\mathop {\lim }\limits_{\xi  \to
\infty } \left( {\frac{{{\sigma _0}}}{P}} \right)\leq \mathop {\lim
}\limits_{\beta \to 1} \left( {\frac{{{\sigma _0}}}{P}} \right)$.
For $\nu_S=0.5$, we have $\gamma=1$ through the definition in Eq.~(\ref{eqGAmma}) and both limits in Eqs.~(\ref{LimitNosubstratSigma}) and~(\ref{LimitXiInfiniSigma})
are equal. Thus, the hoop stress in a very stiff ring lying on a soft
incompressible substrate is the same as the hoop stress in a ring
with no substrate, which serves as a verification of the rationale thus far.

\subsubsection{The instability: order 1 in $\varepsilon$ }

At order 1 in $\varepsilon$, the Euler-Lagrange Eqs.~(\ref{eqEuler})
write
\begin{subequations}\label{eqOrder1}
\begin{align}
\left( {{a_1}p + \widetilde {{a_1}}} \right)A + \left( {{b_1}p +
\widetilde {{b_1}}} \right)B &= 0,\\
\left( {{a_2}p + \widetilde {{a_2}}} \right)A + \left( {{b_2}p +
\widetilde {{b_2}}} \right)B &= 0,
\end{align}
\end{subequations}
where $a_i$, $\widetilde {{a_i}}$, $b_i$ and $\widetilde {{b_i}}$
are functions of the azimuthal wavenumber $m$, the stiffness ratio
$\xi$, the cavity size $\beta$ and the substrate Poisson's ratio
$\nu_S$, reported in full in \ref{Appendix Linear analysis of
stability}. The linear system in Eq.~(\ref{eqOrder1}) has a
nontrivial solution when its determinant vanishes, which for a given
value of $m$ occurs at the dimensionless pressure
\begin{equation}\label{eqPm}
{p_m} = \frac{{{P_m}}}{{\overline {{E_F}} }} = \frac{{ - \widetilde
{{a_1}}\widetilde {{b_2}} + \widetilde {{b_1}}\widetilde
{{a_2}}}}{{{a_1}\widetilde {{b_2}} + \widetilde {{a_1}}{b_2} -
{b_1}\widetilde {{a_2}} - \widetilde {{b_1}}{a_2}}}.
\end{equation}
The critical azimuthal wavenumber and the dimensionless critical
pressure can now be obtained from Eq.~(\ref{eqPm}) by minimizing
over all possible values of $m$
\begin{equation}\label{MCandPC}
\left[ {{m_c},{p_c}} \right] = \mathop {\min }\limits_{\scriptstyle
m= 2, 3,...{\rm{ }}\hfill} \left( {{p_m}} \right).
\end{equation}
For the general case, this minimization must be performed numerically.
However, in the absence of the substrate, we obtain that $m_c=2$ and Eq.~(\ref{eqPm})
simplifies to
\begin{equation}\label{CritStressNoSubtrate}
\mathop {\lim }\limits_{\beta  \to 1} \left( {{p_c}} \right) =
\mathop {\lim }\limits_{\beta  \to 1} \left( {{p_2}} \right) =  -
\frac{1}{{4 - \chi }}{h^3} - {\mkern 1mu} \frac{4}{{{{\left( {4 -
\chi } \right)}^2}}}{h^5} + O\left( {{h^7}} \right).
\end{equation}
The term of order $h^3$ corresponds to the classical dimensionless
critical pressure for a ring with no substrate
\citep{Levy84,Carrier47,Timoshenko61,Combescure81}. The term in
$h^5$ is a correction also reported by \cite{Atanackovic96}. In the
limit of a ring much stiffer than the substrate, we also have $m_c=2$,
with
\begin{equation}\label{CriticalPressureStiff}
\mathop {\lim }\limits_{\xi  \to \infty } \left( {{p_c}} \right) =
\mathop {\lim }\limits_{\xi  \to \infty } \left( {{p_2}} \right) = -
\frac{1}{2}{\mkern 1mu} \frac{{\left( { - 1 + 2{\mkern 1mu} {\nu _S}
- {\beta ^2}} \right){h^3}}}{{{\beta ^2}\left( { - 1 + {\nu _S}}
\right)\left( {4 - \chi } \right)}} - {\mkern 1mu} \frac{{2\left( {
- 1 + 2{\mkern 1mu} {\nu _S} - {\beta ^2}} \right){h^5}}}{{{\beta
^2}\left( { - 1 + {\nu _S}} \right){{\left( {4 - \chi }
\right)}^2}}} + O\left( {{h^7}} \right),
\end{equation}
so that $\mathop {\lim }\limits_{\xi  \to \infty } \left( {{p_c}}
\right) \leq\mathop {\lim }\limits_{\beta  \to 1} \left( {{p_c}}
\right)<0$. Again, for $\nu_S=0.5$, we observe that the limits equal, corroborating the physical intuition that as
the stiffness ratio $\xi$ increases, the effect of the substrate becomes negligible.

\subsection{Critical stress, wavelength and comparison to wrinkling
on a planar substrate}

To determine the effect of curvature on the instability, we compare
the critical hoop stress, $\sigma_c$, in the ring, given by
Eq.~(\ref{eqV0}) with $p=p_c$, as well as the wavelength $\lambda_c= 2\pi
R/m_c$ of the wrinkling mode, against their counterparts for an initially planar film on a plane substrate of infinite thickness
\citep{Chen04}
\begin{subequations}\label{Eq_Sig_lambda_plane}
\begin{align}
{\sigma _{Plane}} &= \frac{{\overline {{E_F}} }}{4}{\left(
{3\frac{{{{\overline {{E_S}} }^*}}}{{\overline {{E_F}} }}}
\right)^{{2 \mathord{\left/
 {\vphantom {2 3}} \right.
 \kern-\nulldelimiterspace} 3}}} = \frac{{\overline {{E_F}} }}{4}{\left( {\frac{{4{{\left( {1 - {\nu _S}} \right)}^2}}}{{3 - 4{\nu _S}}}\frac{3}{\xi }} \right)^{{2 \mathord{\left/
 {\vphantom {2 3}} \right.
 \kern-\nulldelimiterspace} 3}}},\\
{\lambda _{Plane}} & = 2\pi H{\left( {\frac{1}{3}\frac{{\overline
{{E_F}} }}{{{{\overline {{E_S}} }^*}}}} \right)^{{1 \mathord{\left/
 {\vphantom {1 3}} \right.
 \kern-\nulldelimiterspace} 3}}} = 2\pi H{\left( {\frac{1}{3}\frac{{3 - 4{\nu _S}}}{{4{{\left( {1 - {\nu _S}} \right)}^2}}}\xi } \right)^{{1 \mathord{\left/
 {\vphantom {1 3}} \right.
 \kern-\nulldelimiterspace} 3}}},
\end{align}
\end{subequations}
where ${\overline {{E_S}} ^*}=4\overline {{E_S}}{{\left( {1 - {\nu
_S}} \right)}^2}/(3 - 4{\nu _S})$ is the effective stiffness of the
substrate. Together, Eqs.~(\ref{Eq_Sig_lambda_plane}) and
(\ref{eqV0}) yield
\begin{subequations} \label{Eq_Comparison_Plane}
\begin{align}
\frac{{{\sigma _c}}}{{{\sigma _{Plane}}}} &= \frac{{4\gamma
{p_c}}}{{ {h + {{{h^3}} \mathord{\left/
 {\vphantom {{{h^3}} {12}}} \right.
 \kern-\nulldelimiterspace} {12}} + {\mkern 1mu} {k_0}} }}{\left( {\frac{{12{{\left( {1 - {\nu _S}} \right)}^2}}}{{3 - 4{\nu _S}}}\frac{1}{\xi }} \right)^{{{ - 2} \mathord{\left/
 {\vphantom {{ - 2} 3}} \right.
 \kern-\nulldelimiterspace} 3}}},\label{Eq_Comparison_Plane_Sigma}\\
\frac{{{\lambda _c}}}{{{\lambda _{Plane}}}} &=
\frac{1}{{h{m_c}}}{\left( {\frac{{3 - 4{\nu _S}}}{{12{{\left( {1 -
{\nu _S}} \right)}^2}}}\xi } \right)^{{{ - 1} \mathord{\left/
 {\vphantom {{ - 1} 3}} \right.
 \kern-\nulldelimiterspace} 3}}},\label{Eq_Comparison_Plane_Lambda}
\end{align}
\end{subequations}
and, in the limit of absence of the substrate, Eq.~(\ref{Eq_Comparison_Plane_Sigma}) reduces to
{\color{black}{
\begin{equation}\label{Eq_Asymp_beta_1}
\mathop {\lim }\limits_{\beta  \to 1} \left( {\frac{{{\sigma
_c}}}{{{\sigma _{Plane}}}}} \right) =\frac{1}{{{\sigma
_{Plane}}}}\mathop {\lim }\limits_{\beta  \to 1} \left( {{\sigma
_c}} \right)=\frac{1}{{{\sigma _{Plane}}}}\mathop {\lim
}\limits_{\beta  \to 1} \left( {{\sigma _2}} \right) = \frac{4}{{4 -
\chi }}{\left( {\frac{{12{{\left( {1 - {\nu _S}}
\right)}^2}}}{{\left( {3 - 4{\mkern 1mu} {\nu _S}} \right)\xi }}}
\right)^{ - {2 \mathord{\left/
 {\vphantom {2 3}} \right.
 \kern-\nulldelimiterspace} 3}}}{h^2} + O\left( {{h^4}} \right),
\end{equation}
where $\nu_S$ is the Poisson's ratio of the substrate in the planar case.}}

In \S~\ref{Section_Comparison_to_wrinkling_on_a_plane}, we shall
make use of Eqs.~(\ref{Eq_Comparison_Plane}) and
Eq.~(\ref{Eq_Asymp_beta_1}) to further quantify the sensitivity of
the stress and wavelength to the curvature of the substrate.

\section{Numerical simulations}
\label{sec:fem}

In \S~\ref{sec:results}, we will contrast  the predictions from
the above analysis with the results of a series of finite element
simulations performed using the commercial package Abaqus, with the
\verb=BUCKLE= analysis, which provides the buckling load and the
corresponding eigenmodes.

As in the analytical study, the cylindrical structure was modeled as
an annulus of a soft substrate with a stiffer thin film adhered to
its exterior, under plane strain conditions.  The bonding between
the substrate and the film is assumed to be perfect, such that both
share nodes. Negative pressure is applied on the interior surface,
to model the effect of the pressure differential between the inner
surface of the annulus (the cavity) and the exterior of the system.
The buckling analysis provides the value of the critical pressure,
$p_c$, as well as the corresponding critical mode. Rigid body
motions are removed by constraining the displacement of two points
on the film.

Both substrate and film are modeled as incompressible linearly elastic
materials, $\nu_S=\nu_F=0.5$.
These results were compared to additional simulations using a
Neo-Hookean model but no difference was observed given the low values of strain involved. The
substrate was modeled using quadrilateral plane strain elements. Due
to the incompressibility, the corresponding hybrid element, CPE4H,
was used. The film was modeled using B21 beam elements. In order to
account for the effect of plane strain, the stiffness of the beam is
defined as $E_F / \left( 1 - \nu_F^2 \right)$.  

All of the results presented were obtained using 1000 elements in
the circumferential direction, and  $150 \left(R - R_0 \right)$
elements in the radial direction. The mesh size was validated with a
convergence analysis. By way of example, differences less than 0.5\%
in critical pressure were obtained when comparing results for a mesh with
twice the elements in each direction, even in the cases of wrinkling
with the shorter wavelengths. The deviations between the two meshes
were, however, larger ($\sim 5\%$) for the case when the critical
buckling mode is a Biot instability, due to the
infinite number of wavelengths associated to the same buckling mode.
However, as stated in \S~\ref{Subsection Presentation}, this mode shall not be
studied in detail as we focus on the wrinkling and global
buckling loads.

{\color{black}{To test the validity of using beam theory to describe the ring in our problem, we also
performed numerical simulations with 2D solid elements (CPE4H, the same used for the
substrate) and found excellent agreement with the simulations using beam elements. However, in order to achieve such agreement, the mesh needs to be greatly refined, resulting in
a significant increase in computational cost. Attempts to use a mesh size similar to that
of our previous simulations showed clear disagreement. Given the excellent agreement
between the two versions - either using CPE4H (with a fine mesh) or B21 for the ring -
as well as the significantly lower computational cost of the B21 elements, we have decided
to use beam elements for the ring in our analysis.}}

\section{Results}
\label{sec:results}

Having introduced our analytical and numerical methods, we now
present the results of a systematic exploration of the mechanical
response of our system for different geometric and material
parameters. Throughout, we provide a direct comparison between
analytical results and numerical simulations, finding good
agreement. A few instances of discrepancy will also be discussed.

For the geometric parameters, we have varied the cavity size,
$\beta=R_0/R$, and the dimensionless thickness, $h = H/R$. Three
representative values were chosen for $\beta$ (depicted in the
insets of Fig.~\ref{FigHoop}a-c): a small cavity, $\beta = 0.2$; a
cavity with size half of the external radius, $\beta = 0.5$; and a
large cavity, $\beta =0.8$.  Moreover, $h$ was varied in the range
$10^{-3}$ to $10^{-1}$. This parameter has two different physical
interpretations. On one hand, for a substrate with given curvature,
\emph{i.e.} fixed $R$, increasing $h$ is equivalent to increasing
the thickness of the ring. On the other hand, for a ring of given
thickness $H$, the value of $h$ decreases with the curvature, $1/R$.

For the material properties, we have considered values for the
stiffness ratio between the film and the substrate,
$\xi=\bar{E}_F/\bar{E}_S$, spanning over five orders of magnitude,
from $10^2$ to $10^7$. The substrate is taken to be incompressible,
$\nu_S = 0.5$, since most of the relevant experiments that have
motivated our study \citep{Bowden98,Huck00,Yu10,Terwagne14} use
nearly incompressible elastomeric substrates. Moreover, it is
important to note that, even if the analytical model has been
derived assuming a general value of $\nu_S$, it is expected to be
less accurate for increasing deviations from incompressibility.
This was analyzed by \cite{Cai11}, who showed that for the wrinkling
of plates deviations when $\nu_S = 0.3$ are of just a few percent.

In the presentation of our results, we first consider the effect of $h$ and $\xi$ on the hoop-stress
and on the critical pressure. Then, we rationalize the transition
from wrinkling to global buckling, shown in Fig. \ref{Fig0}. We
finally compare the critical stress and wavelength of the
wrinkling mode to their planar substrate counterparts and discuss
the effect of curvature.

\subsection{Hoop stress prior to wrinkling}

\begin{figure}[h!]
    \begin{center}
\includegraphics[width=0.75\columnwidth]{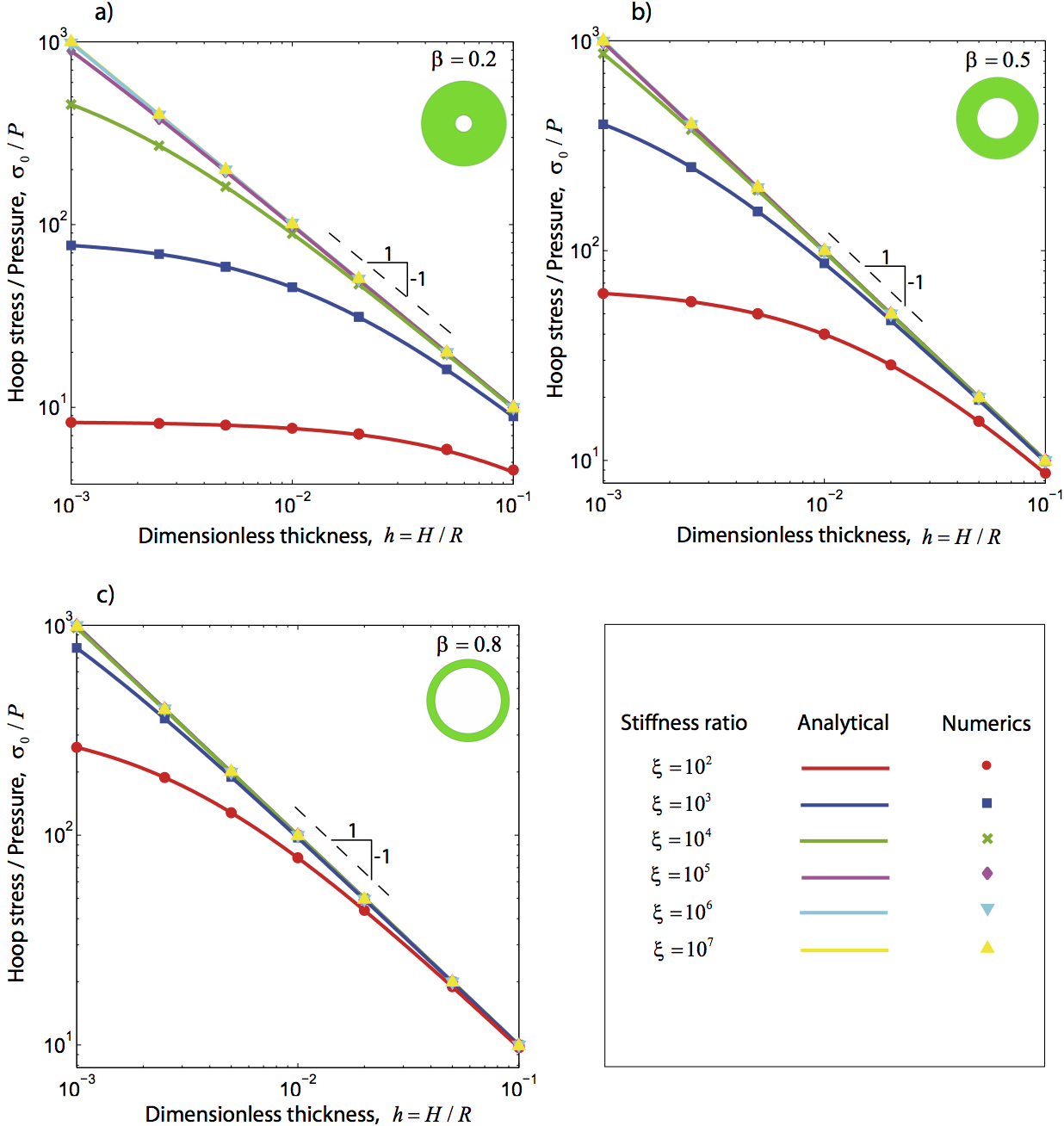}
\end{center}
\caption{Pre-instability hoop stress, $\sigma_0$, in the ring,
normalized by the pressure, $P$, as a function of the dimensionless
ring thickness, $h=H/R$. Cavity sizes are: (a) $\beta=0.2$, (b)
$\beta=0.5$ and (c) $\beta=0.8$.  The Poisson's ratio of the substrate
is $\nu_S=0.5$. Analytical predictions are given by Eq.~(\ref{eqV0})
as solid lines and FEM results are plotted as data points. The
legend (bottom right) is common to all three plots.}\label{FigHoop}
\end{figure}

In Fig.~\ref{FigHoop}, we plot the hoop stress in the ring,
normalized by the pressure, $\sigma_0/P$, as a function of the
dimensionless thickness, $h=H/R$. We find that the hoop stress
decreases monotonically with the dimensionless thickness, and
increases with both the stiffness ratio and the cavity size. When
the ring is much stiffer than the substrate (i.e.
$\xi\rightarrow\infty$), the hoop stress scales as
$\sigma_0/P\sim(H/R)^{-1}$, with a prefactor given by
Eq.~(\ref{LimitXiInfiniSigma}).
For a given thickness of the ring and a given cavity size, the hoop stress in the ring decreases as the curvature
of the substrate increases. This observation is consistent with the classic result
for the hoop stress, $\sigma_0=PR/H$, for a depressurized thin-walled cylindrical pressure
vessel. This result can be recovered from Eq.~(\ref{eqV0}) by
taking $\gamma=1$, $k_0=0$ and performing a Taylor expansion in $h$,
about 0.

\subsection{Critical pressure}
\label{sec:critical_pressure}

\begin{figure}[h!]
    \begin{center}
\includegraphics[width=0.75\columnwidth]{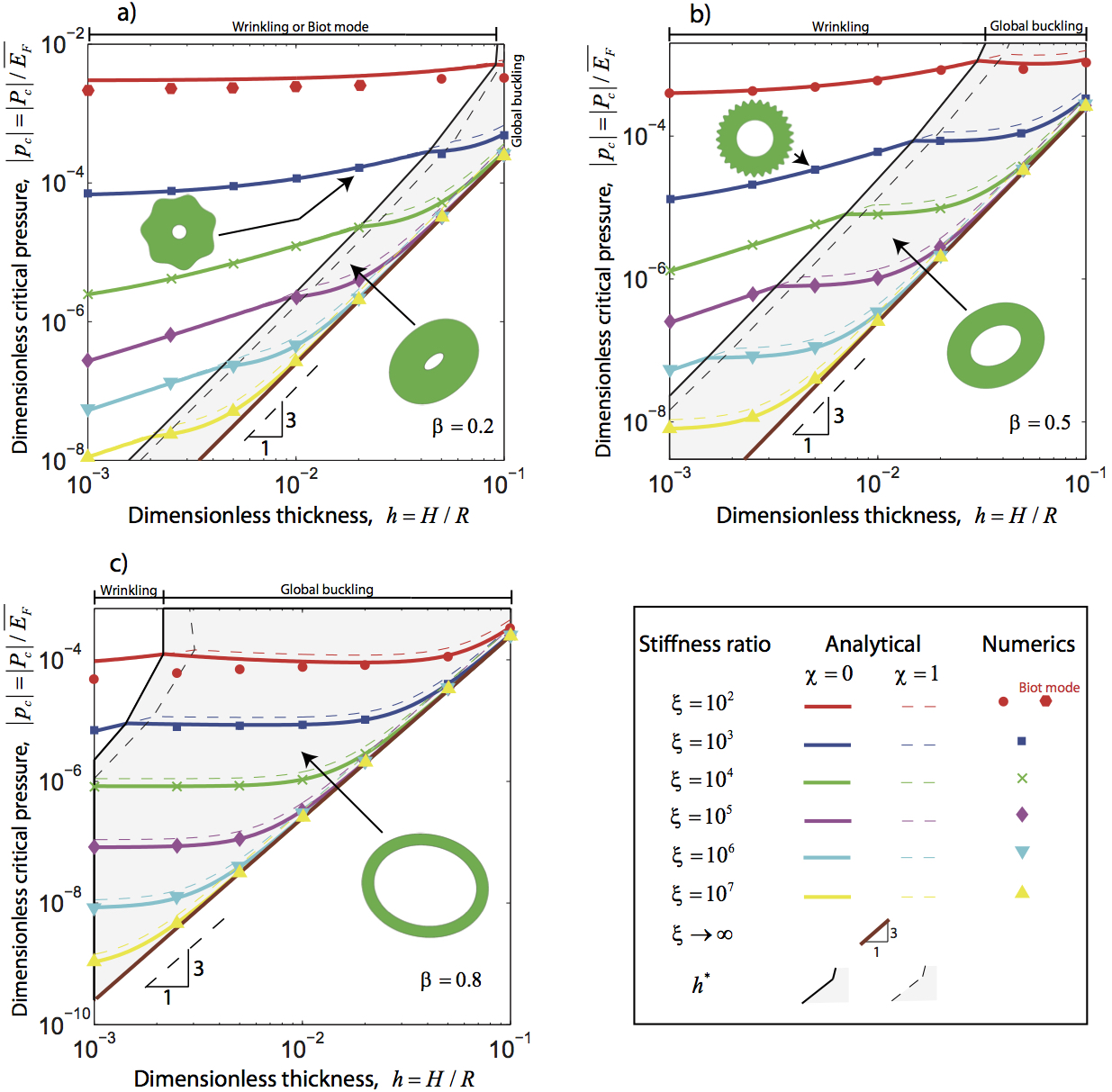}
\end{center}
\caption{Dimensionless critical pressure
$|p_c|=|P_c|/\overline{E_F}$ versus the dimensionless ring thickness
$h=H/R$, for several stiffness ratios
$\xi=\overline{E_F}/\overline{E_S}$ and cavity sizes $\beta$: (a)
$\beta=0.2$, (b) $\beta=0.5$ and (c) $\beta=0.8$. Theory (lines) is given by
Eq.~(\ref{MCandPC}). The Poisson's ratio of the substrate is
$\nu_S=0.5$. The legend (bottom right) is common to all three
plots.}\label{CriticalPressure}
\end{figure}

In Fig.~\ref{CriticalPressure}, we plot the dimensionless critical
pressure, $|p_c|=|P_c|/\overline{E_F}$, as a function of the
dimensionless thickness, $h$, and observe two different regimes. For
low values of $h$, the ring wrinkles with an azimuthal wavenumber
$m_c\gg2$, with a critical pressure $|p_c|$ that increases with $h$.
When $h$ reaches a threshold value, $h^{*}$, the wavenumber
decreases suddenly to $m_c=2$; the nature of the instability changes from a
wrinkling to a global buckling mode. In this regime,
there is a clear asymptote as $\xi \to \infty$, given by
Eq.~(\ref{CriticalPressureStiff}), which corresponds to the buckling
of a ring with no substrate.

Depending on the model used for the reaction force of the substrate
(constant direction force, $\chi=1$, or follower force, $\chi=0$,
see \S~\ref{Subsection Substrate}), Eq.~(\ref{eqForceOnFilm2models})
yields two different analytical predictions, shown in
Fig.~\ref{CriticalPressure} as a solid line for $\chi = 0$, and a
dashed line for $\chi = 1$. We note that the agreement between FEM
simulations and analytical predictions is superior for $\chi=0$, in
particular for the global mode. In other words, it is better to
model the reaction force of the substrate as a pressure field which
remains normal to the ring center-line $C$ as it deforms, than as a
force with constant direction. Thus, from now on, all analytical
results will be only presented for $\chi=0$.

Despite the overall good agreement between the analytical and numerical
results, there are noticeable discrepancies in the global modes for low values of the stiffness ratio.  {\color{black}{ The reason is that, for $\xi = 10^2$, the strains in the substrate can become significant, such that the assumptions for our linear theory are no longer valid (see \ref{Appendix Response of the substrate}). A particularly extreme case is the appearance of Biot modes for $\beta = 0.2$ and $\xi = 10^2$. These modes, although possible, only occur in a small region of our parameter space and, as mentioned above, are beyond the scope of this work. }}

In short, our model exhibits limitations if the ring and the substrate have comparable stiffness, or when there is a Biot mode (small cavity). Apart from this extreme combination of parameters, rarely observed in experimental configurations, the model performs successfully.

\subsection{Phase diagram}
\label{sec:phase_diagram}

We proceed by  focusing on the transition from wrinkling to global
buckling, towards first constructing a phase diagram  in the
$(\xi,h)$ parameter space and then quantifying the dependence of the
boundary, $h^{*}(\xi)$, between the two modes on the size of
the cavity, $\beta$. Given that we do not have a closed form
expression for $h^{*}$, we use a numerical method which tracks any
jump from $m_c=2$ to $m_c^*>2$ in Eq.~(\ref{MCandPC}), when
 $h$ is decreased.

\begin{figure}[h!]
\centering
\includegraphics[width=0.75\columnwidth]{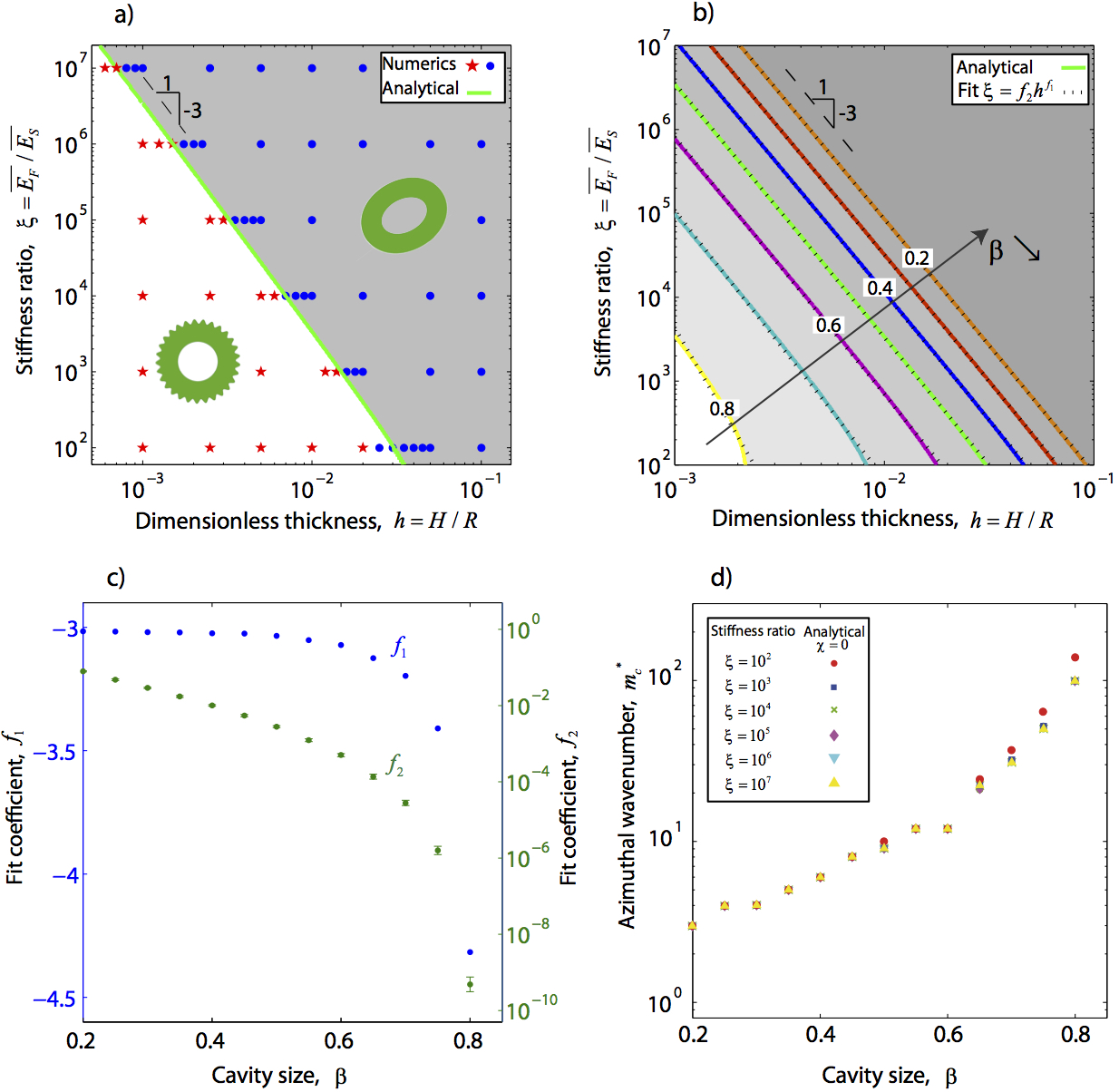}
\caption{(a) Phase diagram in the $(\xi,h)$ parameter space, showing
the transition boundary, $h^{*}$, from wrinkling to global buckling,
for $\beta=0.5$. (b) Phase diagram for $\beta=0.2$ to $\beta=0.8$.
Solid lines are analytical predictions and dotted lines correspond
to the numerical fit $\xi = f_2 h^{f_1}$. (c) Fitting coefficients,
$f_1$ and $f_2$, as functions of $\beta$. (d) Azimuthal wavenumber
of the wrinkling mode for $h=h^{*}$, as a function of $\beta$. The
Poisson's ratio of the substrate is $\nu_S=0.5$.}
\label{fig:phase_diagram}
\end{figure}

In Fig. \ref{fig:phase_diagram}(a), we plot $h^{*}$ as a function of
$\xi$, for $\beta=0.5$. The boundary between modes is consistent
with a power-law, $h^{*}\sim\xi^{-3}$, which divides the phase
diagram into a wrinkling domain ($h<h^{*}$) and a global buckling
domain ($h>h^{*}$, shaded region). Again, there is good agreement
between numerical and analytical solutions.

In Fig. \ref{fig:phase_diagram}(b), we extend the phase diagram to
cavity sizes from $\beta=0.2$ to $\beta=0.8$. We predict that the
power-law with exponent $-3$, mentioned above for $\beta=0.5$, is
still valid as $\beta$ increases, even if there are some deviations
towards the higher values. To quantify the appropriateness of the
$-3$ power-law, we fit a curve of the form $\xi=f_2 h^{f_1}$ to the
analytically calculated boundaries. In Fig.
\ref{fig:phase_diagram}(c) we plot the fitting parameters $f_1$ and
$f_2$ as a function of $\beta$. We find that the exponent is
$f_1\approx -3$ for $\beta < 0.7$, and then decreases rapidly to a
value of $f_1\approx -4.25$ for $\beta = 0.8$. The decrease of the
prefactor $f_2$ is more pronounced and reflects the fact that
$h^{*}$ decreases as $\beta$ increases.

The evolution of $m_c^*$ (i.e. azimuthal wavenumber of the wrinkling
mode for $h=h^{*}$) as a function of $\beta$ and $\xi$ is presented
in Fig.~\ref{fig:phase_diagram}(d). We observe that
$m_c^*$ increases with $\beta$, while it is nearly insensitive to
$\xi$. For example, for $\beta=0.5$, we find that $m_c^*=9$ for
$\xi>10^2$ and $m_c^*=10$ for $\xi=10^2$. Predictions for $\xi>10^2$
are in relatively good agreement with finite element simulations,
which have shown that $m_c^*=10$. For $\xi=10^2$, FEM simulations
yield $m_c^*=15$, pointing out, once again,  the limitation of our
analytical approach in the limit when the stiffness of the ring and
the substrate become comparable.


\subsection{Comparison of wrinkling in our curved system with that on infinite planar substrate}
\label{Section_Comparison_to_wrinkling_on_a_plane}

Thus far, we have shown that the ring may wrinkle or
buckle globally, depending on the curvature and stiffness of the substrate.
We now focus on the wrinkling mode of the ring, with the aim of
comparing the critical stress and wavelength of our curved system to their counterparts for a infinite planar substrate. We shall center our discussion of this comparison for $\beta=0.5$. In  \ref{Appendix Cavity size}, we report the results for $\beta=0.2$ and $\beta=0.8$, which are qualitatively similar.

\begin{figure}[h!]
    \begin{center}
\includegraphics[width=1.\columnwidth]{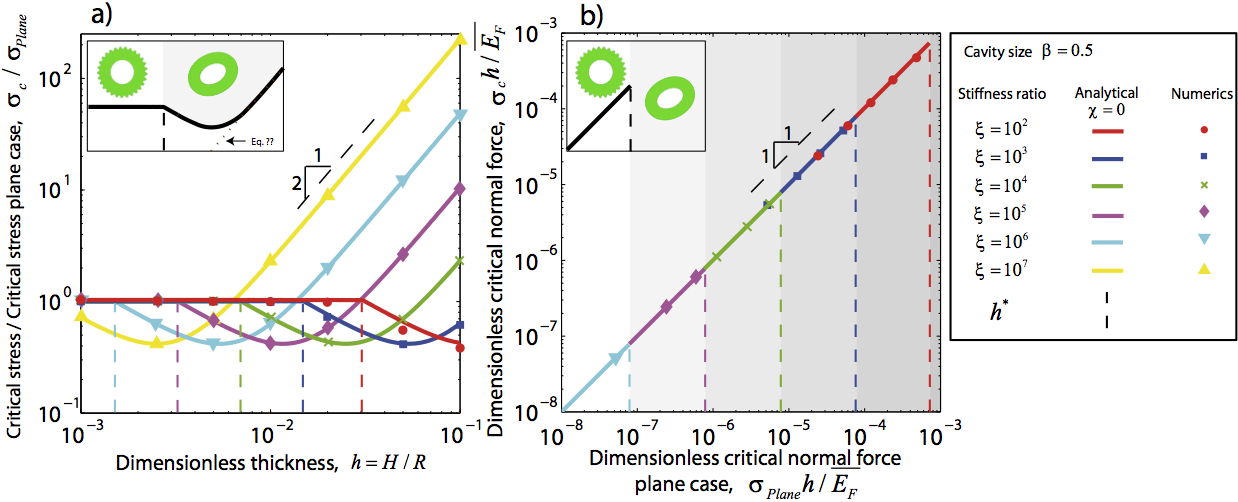}
\end{center}
\caption{(a) Critical stress $\sigma_c$ normalized by its planar
substrate counterpart, $\sigma_{Plane}$, as a function of the
dimensionless ring thickness $h=H/R$. (b) Dimensionless critical
normal force versus planar substrate counterpart. Analytical
predictions (solid lines) are given by
Eq.~(\ref{Eq_Comparison_Plane_Sigma}) and FEM results are shown as
data points. Insets are  sketches of the wrinkling and global
buckling modes. The cavity size is $\beta=0.5$ and the Poisson's
ratio of the substrate is $\nu_S=0.5$. The legend is common to both
plots.}\label{ComparisonPlaneStress}
\end{figure}

In Fig.~\ref{ComparisonPlaneStress}(a), we plot
$\sigma_c/\sigma_{Plane}$ given by
Eq.~(\ref{Eq_Comparison_Plane_Sigma}), as a function of $h$. For
$h<h^{*}$ (wrinkling domain), we find that
$\sigma_c/\sigma_{Plane}\approx1$, in agreement with the FEM
simulations. To further quantify how close to unity is this ratio,
in Fig. \ref{ComparisonPlaneStress}(b) we plot the dimensionless
critical normal force $\sigma_c h/\overline{E_F}$ as a function of
the planar result, $\sigma_{Plane} h/\overline{E_F}$. We obtain a
line with unit slope, indicating that the substrate curvature has no
significative effect on the critical stress for wrinkling. From this
observation, and using Eq.~(\ref{eqV0}) with
$\sigma_0=\sigma_{Plane}$, we write the following approximation for
the dimensionless critical pressure,
\begin{equation}
|p_c| = \frac{{h + {k_0}}}{\gamma }\frac{{{\sigma
_{Plane}}}}{{\overline {{E_F}} }} + O\left( {{h^3}} \right),
\end{equation}
with $k_0$ and $\gamma$ given by Eq.~(\ref{eqDimlessStiff}).
Finally, back to Fig. \ref{ComparisonPlaneStress}(a), for $h>h^{*}$
(global buckling domain), we find that $\sigma_c/\sigma_{Plane}$
first decreases with $h$, then reaches a local minimum and
eventually increases as a power-law with slope 2,  towards the
asymptotic limit given by Eq.~(\ref{Eq_Asymp_beta_1}). The evolution
of $\sigma_c/\sigma_{Plane}$ in the global buckling domain is also
reproduced well by FEM simulations.

\begin{figure}[h!]
    \begin{center}
\includegraphics[width=0.75\columnwidth]{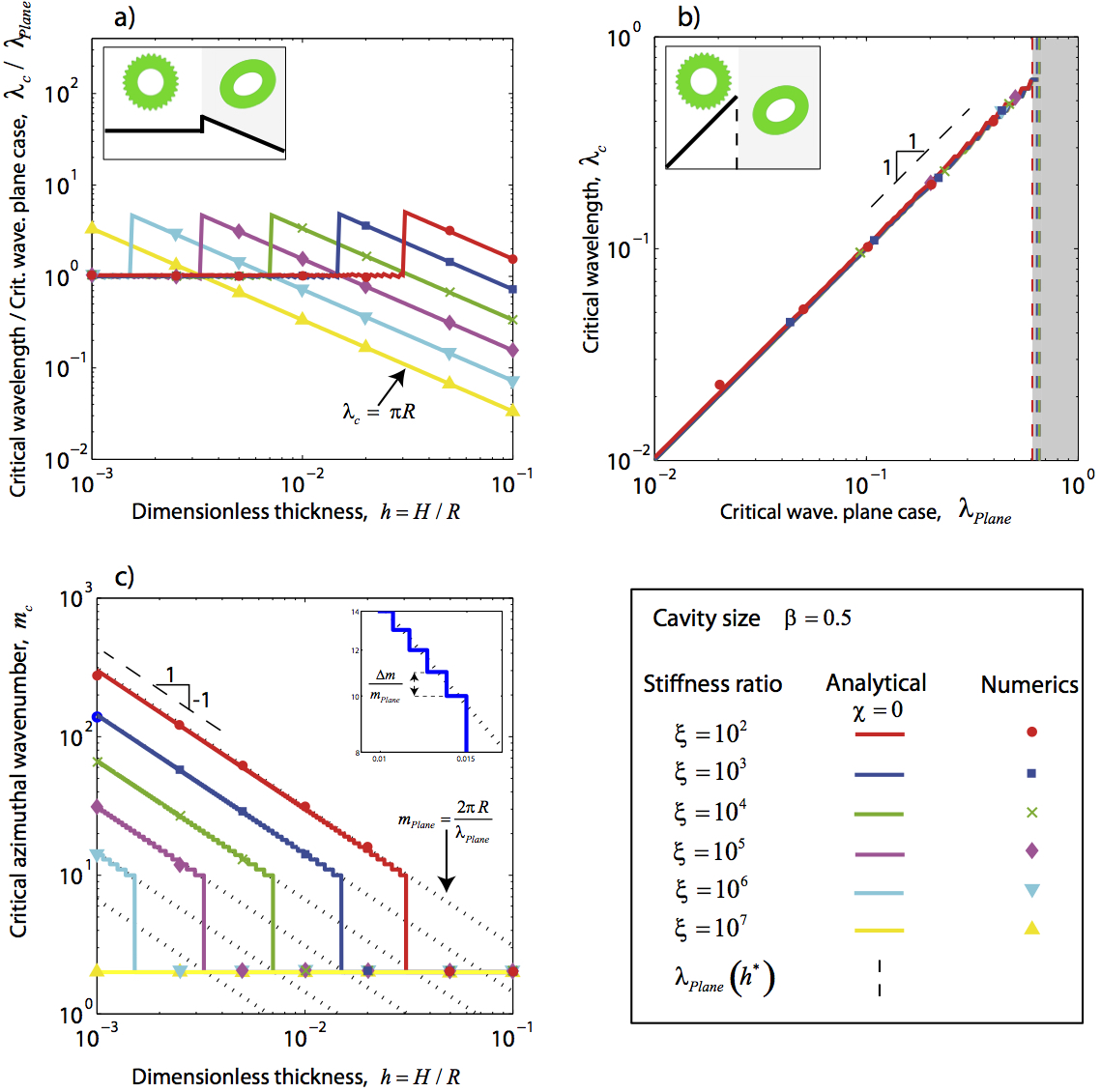}
\end{center}
\caption{(a) Critical wavelength, $\lambda_c$, normalized by its
planar substrate counterpart, $\lambda_{Plane}$, as a function of
the dimensionless ring thickness $h=H/R$. Analytical predictions
(solid lines) are given by Eq.~(\ref{Eq_Comparison_Plane_Lambda}).
(b) Zoom of the wrinkling domain by plotting $\lambda_c$ versus
$\lambda_{Plane}$. (c) Azimuthal wavenumber $m_c$ of the wrinkling
mode as a function of $h$. Analytical predictions (solid lines) are
given by Eq.~(\ref{MCandPC}). Dotted lines show $m_c$ for the planar
substrate case. Insets on (a) and (b) are  sketches of the wrinkling
and global buckling modes. Inset on (c) is a zoom in showing the
difference in $m_c$ for the curved and planar substrates. The cavity
size is $\beta=0.5$ and the Poisson's ratio of the substrate is
$\nu_S=0.5$. FEM results shown as data points. The legend (bottom
right) is common to all three plots.} \label{ComparisonPlaneWave}
\end{figure}

We now investigate the effect of curvature on the wavelength of the
instability mode. In Fig. \ref{ComparisonPlaneWave}(a), we plot
$\lambda_c/\lambda_{Plane}$, given by
Eq.~(\ref{Eq_Comparison_Plane_Lambda}), as a function of $h$.
Focusing on the wrinkling domain, we find that
$\lambda_c/\lambda_{Plane}\approx1$, in agreement with FEM
simulations. In Fig. \ref{ComparisonPlaneWave}(b), once again, we quantify how close to unity this ratio is by plotting  the critical wavelength $\lambda_c$ as a function of the planar result, $\lambda_{Plane}$. {\color{black}{To first
approximation, noting the large dynamic range (at least two orders of magnitude) in both axes of the plot}}, we find a line with unit slope that passes through the origin, suggesting that the substrate curvature has no significative effect on the wavelength of
the wrinkling mode. The only deviations arise from the discrete nature of the wavenumber,
since the geometry of the ring enforces $2\pi$ periodic wrinkling
modes. To highlight this phenomenon, in Fig.
\ref{ComparisonPlaneWave}(c), we plot the critical azimuthal wavenumber
$m_c$, given by Eq.~(\ref{MCandPC}), as a function of $h$ and we
superimpose  the planar substrate result (dashed line),
$m_{Plane}=2\pi R/\lambda_{Plane}$. We find that $m_c$ is a
decreasing stair function of $h$. The deviation in wavelength
between the curved and planar cases scales as
$(|\lambda_c-\lambda_{Plane}|)/\lambda_{Plane}\sim 1/m$, which is
maximum for $m_c^*$. As  shown previously in Fig.
\ref{fig:phase_diagram}(d), $m_c^*$ increases with $\beta$, hence
the deviation $(|\lambda_c-\lambda_{Plane}|)/\lambda_{Plane}$ is
maximum for a small cavity size. By way of example, for $\beta=0.5$,
Fig. \ref{fig:phase_diagram}(d) indicates that $m_c^*=9$, leading
to a maximum deviation
$(|\lambda_c-\lambda_{Plane}|)/\lambda_{Plane}\approx 10\%$, that
rapidly decreases as $m$ increases.

\section{Conclusion}
\label{sec:discussion}

We have considered the two-dimensional problem of a ring bound to an
elastic substrate which contains a cavity that is depressurized. An energy formulation was used to derive the Euler-Lagrange
equations that govern the equilibrium of the ring, and solved them
via an asymptotic expansion. As an improvement to previous results
in the literature, our analytical approach accounts for the effect
of curvature in modeling the reaction force of the substrate. These analytical
results were compared with numerical simulations.

We  first studied the principal solution, obtaining an expression
for the hoop stress in the ring as a function of the applied
pressure. We  then performed a stability analysis of the problem to
determine the critical pressure, $P_c$, and the corresponding
instability mode. Depending on the dimensionless thickness and
stiffness ratio ($h$ and $\xi$) we have identified two different
regimes: local wrinkling of the ring, and global buckling of the
structure. The boundary between both regions of instability was
described via a detailed phase diagram,  which quantifies the value
of $h$ and $\xi$ at which the transition between instabilities
occurs and takes into account the cavity size, $\beta$. Our results
can be used as a design guideline to target a desired mode. Finally,
we have shown that the critical stress for wrinkling and the resulting wavelength do not depend
significantly on the curvature of the substrate. However, curvature
imposes a discretization of the wrinkling wavelength due to the
periodic closing conditions of the ring.

Our study focused on a 2D curved system which exhibits instability
modes analogous to the cylindrical pattern found for uniaxial
compression of a film on a flat infinite substrate. Considering more
complex loading conditions (\emph{e.g.} also introducing axial
loading) or shells with non-zero Gaussian curvature, should lead to
more complex patterns that deserve to be investigated further.
Having validated the FEM analysis, as well as carefully considering
the elastic response of the curved substrate, extending the study to
these other scenarios and addressing the issue of pattern selection
is an exciting avenue for future research.

\section*{Acknowledgements}

D.T. thanks the Belgian American Education Foundation (B.A.E.F.), the Fulbright Program and the Wallonie-Bruxelles International Excellence Grant WBI World.
M.B. thanks the Fulbright Program.
P.M.R. is grateful to financial support from the National Science Foundation, CMMI-1351449 (CAREER) and Saint-Gobain.

\newpage
\appendix

{\color{black}{

\section{Derivative terms in Euler-Lagrange equation}\label{Appendix Euler Lagrange}

The terms in the Euler-Lagrange equation for the equilibrium of the film, Eq.~(\ref{eqEuler2}), are

\begin{subequations}
\begin{align}
\frac{{\partial {\overline{\mathcal{E}}}}}{{\partial
v}}&=-\frac{h^2}{12} \left( -2{v}^{2}+ \left( 2-4{\it v''} \right)
v+2{\it v''}-{{\it v'}}^{2}+2u{\it u''}-4{\it u'}{\it v''}-2{\it
v'}{\it u''}+2{{\it u'}}^{2}+{u}^{2} \right) \left( {\it
v''}+v-\frac{1}{2}
 \right)\nonumber\\
 &+\frac{1}{2} \left( {v}^{2}+ \left( 2+2{\it u'} \right) v+
{{\it u'}}^{2}+2{\it u'}+ \left( u-{\it v'} \right) ^{2} \right)
 \left( 1+v+{\it u'} \right)\\
\frac{{\partial {\overline{\mathcal{E}}}}}{{\partial
u}}&=\frac{h^2}{24} \left( -2{v}^{2}+ \left( 2-4{\it v''} \right)
v+2{\it v'' }-{{\it v'}}^{2}+2u{\it u''}-4{\it u'}{\it v''}-2{\it
v'}{ \it u''}+2{{\it u'}}^{2}+{u}^{2} \right) \left( u+{\it u''}
 \right)\nonumber\\
 &+\frac{1}{2} \left( {u}^{2}-2u{\it v'}+{{\it v'}}^{2}+
 \left( v+2+{\it u'} \right)  \left( v+{\it u'} \right)  \right)
 \left( u-{\it v'} \right)\\
{\left( {\frac{{\partial {\overline{\mathcal{E}}}}}{{\partial
v^{'}}}} \right)^{'}}&=-\frac{h^2}{12} \left(  \left( -2v+1-3{\it
v''}-{\it u'''} \right) {\it v' }+ \left( 3{\it u''}-2{\it v'''}+u
\right) {\it u'}-3{\it u''} {\it v''}+ \left( 1-2v \right) {\it
v'''}+u{\it u'''} \right)
 \left( {\it v'}+{\it u''} \right)\nonumber\\
&-\frac{h^2}{24} \left(  \left( -4v +2-4{\it u'} \right) {\it
v''}-2{v}^{2}+2v-2{\it v'}{\it u'' }-{{\it v'}}^{2}+2u{\it
u''}+2{{\it u'}}^{2}+{u}^{2} \right)\left( {\it v''}+{\it u'''}
\right)\nonumber\\
&- \left( u-{\it v'} \right)
 \left(  \left( v+1+{\it v''} \right) {\it v'}+ \left( u+{\it u''}
 \right) {\it u'}+ \left( 1+v \right) {\it u''}-u{\it v''} \right)\nonumber\\
 &-\frac{1}{
2} \left( {\it u'}-{\it v''} \right)  \left( {{\it u'}}^{2}+ \left(
2 v+2 \right) {\it u'}+{v}^{2}+2v+ \left( u-{\it v'} \right) ^{2}
 \right)\\
{\left( {\frac{{\partial {\overline{\mathcal{E}}}}}{{\partial
u^{'}}}} \right)^{'}}&=\frac{h^2}{6} \left( {\it u'}-{\it v''}
\right) \left(  \left( -2v+1-3{ \it v''}-{\it u'''} \right) {\it
v'}+ \left( 3{\it u''}-2\,{\it v''' }+u \right) {\it u'}-3{\it
u''}{\it v''}+ \left( 1-2v \right) { \it v'''}+u{\it u'''}
\right)\nonumber\\
&+\frac{h^2}{12} \left( \left( 2u-2{ \it v'} \right) {\it u''}+
\left( -4v+2-4\,{\it u'} \right) {\it v'' }-2{v}^{2}+2v+2{{\it
u'}}^{2}-{{\it v'}}^{2}+{u}^{2} \right)
 \left( -{\it v'''}+{\it u''} \right)\nonumber\\
&+ \left(  \left( u+{\it u''} \right) {\it u'}+ \left( {\it v'}+{\it
u''} \right) v+{\it u''}+
 \left( 1+{\it v''} \right) {\it v'}-u{\it v''} \right)  \left( 1+v+{
\it u'} \right)\nonumber\\
&+\frac{1}{2} \left( {\it v'}+{\it u''} \right) \left( {{ \it
u'}}^{2}+ \left( 2v+2 \right) {\it u'}+{v}^{2}+2v+ \left( u-{ \it
v'} \right) ^{2} \right)\\
{\left( {\frac{{\partial {\overline{\mathcal{E}}}}}{{\partial
v^{''}}}} \right)^{''}}&=-\frac{h^2}{6} \left( v-\frac{1}{2}+{\it
u'} \right) \left( -3{{\it v''}}^{2}+
 \left( 1-2v-4{\it u'''} \right) {\it v''}+{{\it u'}}^{2}+ \left(
-2{\it v''''}+4{\it u'''} \right) {\it u'}+ \left( 1-2v \right) {\it
v''''}\right)\nonumber\\
&-\frac{h^2}{6} \left( v-\frac{1}{2}+{\it u'} \right)\left(3{{\it
u''}}^{2}+ \left( -5{\it v'''}+u \right) {\it u''}-2{{\it v'}}^{2}+
\left( -5{\it v'''}-{\it u''''} \right) { \it
v'}+u{\it u''''} \right)\nonumber\\
&-\frac{h^2}{3} \left( \left( -2v+1-3{ \it v''}-{\it u'''} \right)
{\it v'}+ \left( -2{\it u'}+1-2v
 \right) {\it v'''}+ \left( 3{\it u''}+u \right) {\it u'}+u{\it u'''
}-3{\it u''}{\it v''} \right) \left( {\it v'}+{\it u''}
 \right)\nonumber\\
 &-\frac{h^2}{12} \left(  \left( -4v+2-4{\it u'} \right) {\it v''}-
2{v}^{2}+2v-2{\it v'}{\it u''}-{{\it v'}}^{2}+2u{\it u''}+2 {{\it
u'}}^{2}+{u}^{2} \right)\left( {\it v''}+{\it u'''}
 \right)\\
{\left( {\frac{{\partial {\overline{\mathcal{E}}}}}{{\partial
u^{''}}}} \right)^{''}}&=\frac{h^2}{12}\left( u-{\it v'} \right)
\left( -3{{\it v''}}^{2}+
 \left( 1-2v-4\,{\it u'''} \right) {\it v''}-2{{\it v'}}^{2}+
 \left( -5{\it v'''}-{\it u''''} \right) {\it v'}+ \left( -2{\it
u'}+1-2v \right) {\it v''''}\right)\nonumber\\
&+\frac{h^2}{12}\left( u-{\it v'} \right)\left(\left( {\it
u''''}+{\it u''} \right) u -5{\it u''}{\it v'''}+3{{\it
u''}}^{2}+4{\it u'}{\it u'''}+{ {\it
u'}}^{2} \right)\nonumber\\
&+\frac{h^2}{6} \left( {\it u'}-{\it v''} \right)
 \left(  \left( -2v+1-3{\it v''}-{\it u'''} \right) {\it v'}+
 \left( 3{\it u''}-2{\it v'''}+u \right) {\it u'}-3{\it u''}{
\it v''}+ \left( 1-2v \right) {\it v'''}+u{\it u'''}
\right)\nonumber\\
&+\frac{h^2}{24} \left(  \left( 2u-2{\it v'} \right) {\it u''}+
\left( -4 v+2-4{\it u'} \right) {\it v''}-2{v}^{2}+2v+2{{\it
u'}}^{2}-{{ \it v'}}^{2}+{u}^{2} \right) \left( -{\it v'''}+{\it
u''} \right)
\end{align}
\end{subequations}
and
\begin{subequations}
\begin{align}
\frac{{\partial {\delta \overline W}}}{{\partial \delta
v}}&=\frac{1}{2}{\frac {2{\it \tau} \left( -1+{\it \chi} \right)
\left( -1+{ \it u'}+v \right)  \left( u-{\it v'} \right) -{\it
\sigma} \left(
 \left( -1+{\it \chi} \right) {{\it v'}}^{2}-2u \left( -1+{\it \chi}
 \right) {\it v'}+2+ \left( -1+{\it \chi} \right) {u}^{2} \right)
 }{h}}\\
\frac{{\partial {\delta \overline W}}}{{\partial \delta
u}}&=\frac{1}{2}{\frac {-2{\it \sigma} \left( -1+{\it u'}+v \right)
\left( u-{ \it v'} \right)  \left( -1+{\it \chi} \right) -{\it \tau}
\left(
 \left( -1+{\it \chi} \right) {{\it v'}}^{2}-2u \left( -1+{\it \chi}
 \right) {\it v'}+2+ \left( -1+{\it \chi} \right) {u}^{2} \right) }{h}}
\end{align}
\end{subequations}

}}

\section{Response of the substrate}\label{Appendix Response of the substrate}

In this appendix we consider the boundary value problem of the
substrate, subjected to the ring displacement
$R\left(v\left(\theta\right),u\left(\theta\right)\right)$, $v$ and
$u$ given by Eq.~(\ref{eqKoiter}), at the interface $r=R$ between
the ring and the substrate and to the pressure $P$ at $r=R_0=\beta
R$. The substrate is assumed to be in a state of plane strain. We note
$\alpha  = 3 - 4{\nu _S}$ and introduce the shear modulus $G =
{{\overline {{E_S}} \left( {1 - {\nu _S}} \right)} \mathord{\left/
 {\vphantom {{\overline {{E_S}} \left( {1 - {\nu _S}} \right)} 2}} \right.
 \kern-\nulldelimiterspace} 2}$. The components of the stress in the substrate
are represented by $\sigma_{rr}$, $\sigma_{r\theta}$, and the
displacement field is $(U_r, U_{\theta})$. The 2D problem of
elasticity is solved by finding an Airy function of the
form~\citep{Michell1899},
\begin{equation}\label{Airy}
\phi_m \left( {r,\theta } \right) = {B_1}{r^2} + {B_2}\ln \left( r
\right) + \varepsilon \left( {{A_1}{r^{m + 2}} + {A_2}{r^{ - m + 2}}
+ {A_3}{r^m} + {A_4}{r^{ - m}}} \right)\sin \left( {m\theta }
\right),
\end{equation}
where $A_i$ and $B_i$ are unknown constants determined by the
boundary conditions
\begin{subequations}\label{Boundary conditions}
\begin{align}
{\sigma _{rr}}\left( {{R_0},\theta } \right) &=  - P,\\
{\sigma _{r\theta }}\left( {{R_0},\theta } \right) &= 0,\\
{U_r}\left( {R,\theta } \right) &= Rv\left( \theta  \right),\\
{U_\theta }\left( {R,\theta } \right) &= Ru\left( \theta  \right),
\end{align}
\end{subequations}
which are assumed to apply at $r=R_0$ and $r=R$. The first two
equations stand for the continuity of the stress at the boundary of the cavity, whereas the last two stand for the continuity of the
displacement at the interface between the substrate and the ring.

The stress and displacement fields resulting from the Airy function
Eq.~(\ref{Airy}) are~\citep{Barber02}
\begin{subequations}\label{Stress field}
\begin{equation}
\begin{split}
{\sigma _{rr}}\left( {r,\theta } \right) = 2{B_1} +
\frac{{{B_2}}}{{{r^2}}} + \left( \begin{array}{l}
 - {A_1}\left( {m + 1} \right)\left( {m - 2} \right){r^m} - {A_2}\left( {m + 2} \right)\left( {m - 1} \right){r^{ - m}}\\
 - {A_3}m\left( {m - 1} \right){r^{m - 2}} - {A_4}m\left( {m + 1} \right){r^{ - m - 2}}
\end{array} \right)\sin \left( {m\theta } \right),
\end{split}
\end{equation}
\begin{equation}
\begin{split}
{\sigma _{r\theta }}\left( {r,\theta } \right) = \left(
\begin{array}{l}
 - {A_1}m\left( {m + 1} \right){r^m} + {A_2}m\left( {m - 1} \right){r^{ - m}}\\
 - {A_3}m\left( {m - 1} \right){r^{m - 2}} + {A_4}m\left( {m + 1} \right){r^{ - m - 2}}
\end{array} \right)\cos \left( {m\theta } \right),
\end{split}
\end{equation}
{\color{black}{
\begin{equation}
\begin{split}
{U_r}\left( {r,\theta } \right) = \frac{1}{{2G }}\left[ {{B_1}\left(
{\alpha  - 1} \right)r -\frac{B_2}{r} +\left(
\begin{array}{l}
{A_1}\left( {\alpha  - m - 1} \right){r^{m + 1}} + {A_2}\left( {\alpha  + m - 1} \right){r^{ - m + 1}}\\
 - {A_3}m{r^{m - 1}} + {A_4}m{r^{ - m - 1}}
\end{array} \right)\sin \left( {m\theta } \right)} \right],
\end{split}
\end{equation}
\begin{equation}
\begin{split}
{U_\theta }\left( {r,\theta } \right) = \frac{1}{{2G }} { \left(
\begin{array}{l}
 - {A_1}\left( {\alpha  + m + 1} \right){r^{m + 1}} + {A_2}\left( {\alpha  - m + 1} \right){r^{ - m + 1}}\\
 - {A_3}m{r^{m - 1}} - {A_4}m{r^{ - m - 1}}
\end{array} \right)\cos \left( {m\theta } \right)}.
\end{split}
\end{equation}
}}
\end{subequations}
Applying the boundary conditions Eq.~(\ref{Boundary conditions}) to
Eq.~(\ref{Stress field}) yields a linear system for $A_i$ and $B_i$,
with solution
\begin{subequations}
\begin{equation}\label{B1}
{B_1} = \frac{1}{{2\left( {{\beta ^2} + 1 - 2{\nu _S}}
\right)}}\left[ {\overline {{E_S}} {\mkern 1mu} {v_0}\left( {1 -
{\nu _S}} \right) - P{\beta ^2}} \right],
\end{equation}
\begin{equation}\label{B2}
{B_2} = \frac{{ - {R^2}{\beta ^2}}}{{{\beta ^2} + 1 - 2{\mkern 1mu}
{\nu _S}}}\left[ {\overline {{E_S}} {\mkern 1mu} {v_0}\left( {1 -
{\nu _S}} \right) + {P}\left( {1 - 2{\mkern 1mu} {\nu _S}} \right)}
\right],
\end{equation}
\begin{equation}\label{A1}
\begin{split}
{A_1} &= \frac{{\left( {m - 1} \right){\beta ^{2{\kern 1pt} m + 2}}
+ \left( {\alpha  - m + 1} \right){\beta ^{2{\kern 1pt} m}} + {\beta
^2}{\mkern 1mu} }}{{2\left( { - {\mkern 1mu} {m^2} + 1}
\right){\beta ^{2{\kern 1pt} m + 2}} + {m^2}{\beta ^{2{\kern 1pt} m
+ 4}} + {\beta ^{4{\kern 1pt} m + 2}}\alpha  + \left( {{m^2} - 1 +
{\alpha ^2}} \right){\beta ^{2{\kern 1pt} m}} + \alpha {\beta
^2}}}G{R^{ - m}} A\\
 &\quad+ \frac{{\left( {m - 1} \right){\beta ^{2{\kern 1pt} m + 2}} +
\left( { - m - \alpha  + 1} \right){\beta ^{2{\kern 1pt} m}} -
{\beta ^2}{\mkern 1mu} }}{{2\left( { - {\mkern 1mu} {m^2} + 1}
\right){\beta ^{2{\kern 1pt} m + 2}} + {m^2}{\beta ^{2{\kern 1pt} m
+ 4}} + {\beta ^{4{\kern 1pt} m + 2}}\alpha  + \left( {{m^2} - 1 +
{\alpha ^2}} \right){\beta ^{2{\kern 1pt} m}} + \alpha {\beta
^2}}}G{R^{ - m}} B,
\end{split}
\end{equation}
\begin{equation}\label{A2}
\begin{split}
{A_2} &= \frac{{ - \left( {m + 1} \right){\beta ^{2{\kern 1pt} m + 2}}+ {\beta ^{4{\kern 1pt} m + 2}}+ {\beta ^{2{\kern 1pt} m}}\left( {\alpha  + m + 1} \right)}}{{2\left( { - {m^2} + 1} \right){\beta ^{2{\kern 1pt} m + 2}} + {m^2}{\beta ^{2{\kern 1pt} m + 4}} + {\beta ^{4{\kern 1pt} m + 2}}\alpha  + \left( {{m^2} - 1 + {\alpha ^2}} \right){\beta ^{2{\kern 1pt} m}} + \alpha {\beta ^2}}}G {\mkern 1mu} {R^m}A\\
 &\quad+ \frac{{\left( {m + 1} \right){\beta ^{2{\kern 1pt} m + 2}}
+ {\beta ^{4{\kern 1pt} m + 2}} + {\beta ^{2{\kern 1pt} m}}\left(
{\alpha  - m - 1} \right)}}{{2\left( { - {\mkern 1mu} {m^2} + 1}
\right){\beta ^{2{\kern 1pt} m + 2}} + {m^2}{\beta ^{2{\kern 1pt} m
+ 4}} + {\beta ^{4{\kern 1pt} m + 2}}\alpha  + \left( {{m^2} - 1 +
{\alpha ^2}} \right){\beta ^{2{\kern 1pt} m}} + \alpha {\beta ^2}}}G
{\mkern 1mu} {R^m}B,
\end{split}
\end{equation}
\begin{equation}\label{A3}
\begin{split}
{A_3} &=  - \frac{{{\beta ^2}\left( {\left( {\alpha  - m + 1} \right)\left( {m + 1} \right){\beta ^{2{\kern 1pt} m}} + {m^2}{\beta ^{2{\kern 1pt} m + 2}} + \alpha  + 1 + m} \right)}}{{\left( {2\left( { - {\mkern 1mu} {m^2} + 1} \right){\beta ^{2{\kern 1pt} m + 2}} + {m^2}{\beta ^{2{\kern 1pt} m + 4}} + {\beta ^{4{\kern 1pt} m + 2}}\alpha  + \left( {{m^2} - 1 + {\alpha ^2}} \right){\beta ^{2{\kern 1pt} m}} + \alpha {\beta ^2}} \right)m}}G {\mkern 1mu} {R^{ - m + 2}}A\\
 &\quad- \frac{{{\beta ^2}\left( {{m^2}{\beta ^{2{\kern 1pt} m + 2}} - 1
+ \left( { - m - \alpha  + 1} \right)\left( {m + 1} \right){\beta
^{2{\kern 1pt} m}} - m + \alpha } \right)}}{{\left( {2\left( { -
{\mkern 1mu} {m^2} + 1} \right){\beta ^{2{\kern 1pt} m + 2}} +
{m^2}{\beta ^{2{\kern 1pt} m + 4}} + {\beta ^{4{\kern 1pt} m +
2}}\alpha  + \left( {{m^2} - 1 + {\alpha ^2}} \right){\beta
^{2{\kern 1pt} m}} + \alpha {\beta ^2}} \right)m}}G {\mkern 1mu}
{R^{ - m + 2}}B,
\end{split}
\end{equation}
\begin{equation}\label{A4}
\begin{split}
{A_4} &=  - \frac{{{\mkern 1mu} \left( { - {m^2}{\beta ^{2{\kern 1pt} m + 2}} + \left( {m - 1} \right)\left( {\alpha  + m + 1} \right){\beta ^{2{\kern 1pt} m}} - \left( {\alpha  - m + 1} \right){\beta ^{4{\kern 1pt} m}}} \right){\beta ^2}}}{{\left( {2\left( { - {\mkern 1mu} {m^2} + 1} \right){\beta ^{2{\kern 1pt} m + 2}} + {m^2}{\beta ^{2{\kern 1pt} m + 4}} + {\beta ^{4{\kern 1pt} m + 2}}\alpha  + \left( {{m^2} - 1 + {\alpha ^2}} \right){\beta ^{2{\kern 1pt} m}} + \alpha {\beta ^2}} \right)m}}G {R^{m + 2}}A\\
 &\quad- \frac{{{\mkern 1mu} \left( {{m^2}{\beta ^{2{\kern 1pt} m + 2}}
+ \left( {m - 1} \right)\left( {\alpha  - m - 1} \right){\beta
^{2{\kern 1pt} m}} - \left( { - m - \alpha  + 1} \right){\beta
^{4{\kern 1pt} m}}} \right){\beta ^2}}}{{\left( {2\left( { - {\mkern
1mu} {m^2} + 1} \right){\beta ^{2{\kern 1pt} m + 2}} + {m^2}{\beta
^{2{\kern 1pt} m + 4}} + {\beta ^{4{\kern 1pt} m + 2}}\alpha  +
\left( {{m^2} - 1 + {\alpha ^2}} \right){\beta ^{2{\kern 1pt} m}} +
\alpha {\beta ^2}} \right)m}}G{R^{m + 2}} B.
\end{split}
\end{equation}
\end{subequations}

Substituting $A_i$ and $B_i$ into Eq.~(\ref{Stress field}) yields
the stress at the interface $r=R$
\begin{subequations}\label{Stress_Stiffness}
\begin{align}
{\sigma _{rr}}\left( {R,\theta } \right) &= {K_0}R{v_0} +
KR\varepsilon A\sin \left( {m\theta
} \right) - {\mkern 1mu} \gamma P,\\
{\sigma _{r\theta}}\left( {R,\theta } \right) &= {\rm M}R\varepsilon
B\cos \left( {m\theta } \right),
\end{align}
\end{subequations}
where
\begin{subequations}
\begin{align}
{K_0} &= \overline {{E_S}} \frac{1}{R}\frac{{\left( {1 - {\nu _S}}
\right)\left( {1 - {\beta ^2}} \right)}}{{1 - 2{\mkern 1mu} {\nu _S}
+ {\beta ^2}}},\\
K &= \overline {{E_S}} \frac{1}{R}\frac{{\left( {1 - {\nu _S}}
\right)}}{2}\left( {{S_A}{\mkern 1mu}  + {S_B}\frac{B}{A}}
\right),\\
{\rm M} &= \overline {{E_S}} \frac{1}{R}{\mkern 1mu} {\mkern 1mu}
\frac{{\left( {1 - {\nu _S}} \right)}}{2}\left( {{S_B}{\mkern 1mu}
\frac{A}{B} + {T_B}{\mkern 1mu} } \right),\\
\gamma  &= \frac{{2{\beta ^2}\left( {1 - {\nu _S}} \right)}}{{1 -
2{\mkern 1mu} {\nu _S} + {\beta ^2}}},
\end{align}
\end{subequations}
and
\begin{subequations}
\begin{equation}
\begin{split}
\psi {S_A} &= 2\left( { - {\mkern 1mu} {m^2}\left( {{\beta ^4} + 2 +
\alpha } \right) + \left( {3 + \alpha } \right)\left( {{m^2} - 1}
\right){\beta ^2} + 2{\mkern 1mu} \left( {\alpha  + 1} \right)}
\right){\beta ^{2{\kern 1pt} m}}\\
&\quad- \left( {\left( {\left( {\alpha + 1} \right)m + \alpha  - 1}
\right){\beta ^{4{\kern 1pt} m}} - \left( {\alpha  + 1} \right)m +
\alpha  - 1} \right){\beta ^2},
\end{split}
\end{equation}
\begin{equation}
\begin{split}
\psi {S_B} &=  - 2{\mkern 1mu} m\left( { - {m^2}\left( {{\beta ^4} +
1} \right) + 2\left( {{\mkern 1mu} {m^2} - 1} \right){\beta ^2} + 1
+ \alpha } \right){\beta ^{2{\kern 1pt} m}}\\
&\quad+ \left( {\left( {\left( {\alpha  - 1} \right)m + \alpha  + 1}
\right){\beta ^{4{\kern 1pt} m}} + \left( {\alpha  - 1} \right)m -
\alpha  - 1} \right){\beta ^2},
\end{split}
\end{equation}
\begin{equation}
\begin{split}
 \psi {T_B} &= 2\left( { - {\mkern
1mu} {m^2}{\beta ^4} + \left( {1 - {\mkern 1mu} \alpha }
\right)\left( {{m^2} - 1} \right){\beta ^2} + {\mkern 1mu} \alpha
{m^2}} \right){\beta ^{2{\kern 1pt} m}}\\
&\quad- \left( {\left( {\left( {\alpha  + 1} \right)m + \alpha  - 1}
\right){\beta ^{4{\kern 1pt} m}} - \left( {\alpha  + 1} \right)m +
\alpha  - 1} \right){\beta ^2},
\end{split}
\end{equation}
\begin{equation}
\begin{split}
\psi  &= \left( {{m^2}{\beta ^4} - \left( {{m^2} - 1} \right)\left(
{2{\beta ^2} - 1} \right) + {\alpha ^2}} \right){\beta ^{2{\kern
1pt} m}} + \alpha {\beta ^2}\left( {1 + {\beta ^{4{\kern 1pt} m}}}
\right).
\end{split}
\end{equation}
\end{subequations}

The dimensionless stress at the interface and the dimensionless
stiffness parameters are obtained from Eq.~(\ref{Stress_Stiffness}),
by dividing it with $\overline {{E_F}}$,
\begin{subequations}
\begin{align}
\frac{{{\sigma}}}{{\overline {{E_F}} }}&={k_0}{v_0} + k\varepsilon A\sin (m\theta ) -\gamma p,\\
\frac{{{\tau}}}{{\overline {{E_F}} }}&= \mu \varepsilon B\sin (m\theta),\\
{k_0} &= \frac{{{K_0}R}}{{\overline {{E_F}} }} = \frac{{\left( {1 -
{\nu _S}} \right)\left( {1 - {\beta ^2}} \right)}}{{1 - 2{\mkern
1mu} {\nu _S} + {\beta ^2}}}\frac{1}{\xi },\\
k &= \frac{{{K}R}}{{\overline {{E_F}} }} =\frac{{\left( {1 - {\nu
_S}} \right)}}{2}\left( {{S_A}{\mkern
1mu}  + {S_B}\frac{B}{A}} \right)\frac{1}{\xi },\\
\mu  &=\frac{{{\rm M}R}}{{\overline {{E_F}} }} = {\mkern 1mu}
{\mkern 1mu} \frac{{\left( {1 - {\nu _S}} \right)}}{2}\left(
{{S_B}{\mkern 1mu} \frac{A}{B} + {T_B}{\mkern 1mu} }
\right)\frac{1}{\xi },
\end{align}
\end{subequations}
as indicated in Eq.~(\ref{eqDimlessStiff}).

We  note that in the case of an inextensible wrinkling mode,
$A/B=m$, the stiffness $K$ simplifies to $\widetilde K$ given by
\begin{equation}\label{Stiffness_Simplifiee}
\begin{split}
\widetilde K &= \overline {{E_S}} \frac{1}{R}\frac{{\left( {1 - {\nu
_S}} \right)}}{2}\left( {{S_A}{\mkern 1mu}  + {S_B}\frac{1}{m}}
\right)\\
&= \overline {{E_S}} \frac{1}{{mR}}\frac{{2{{\left( {1 - {\nu _S}}
\right)}^2}\left( {{\beta ^{ - 2{\kern 1pt} m}} - {\beta ^{2{\kern
1pt} m}} + 2{\mkern 1mu} m\left( {1 - {\beta ^{ - 2}}} \right)}
\right)\left( {{m^2} - 1} \right)}}{{{{\left( {{\beta ^{ - 1}} -
\beta } \right)}^2}\left( {{m^2} - 1} \right) + {{\left( {{\beta ^{
- m}} - {\beta ^m}} \right)}^2}\left( {3 - 4{\nu _S}} \right) +
{{\left( {\frac{{3 - 4{\nu _S}}}{\beta } + \beta } \right)}^2}}},
\end{split}
\end{equation}
which, for a substrate with no cavity, leads to the limiting case
\begin{equation}
\mathop {\lim }\limits_{\beta  \to 0} \left( {\widetilde K} \right)
= \overline {{E_S}} \frac{1}{{mR}}\frac{{2{{\left( {1 - {\nu _S}}
\right)}^2}\left( {{m^2} - 1} \right)}}{{3 - 4{\nu _S}}},
\end{equation}
and, for an infinite plane substrate, yields


\begin{equation}\label{Stiffness_Allen}
\mathop {\lim }\limits_{R \to \infty } \left( {\mathop {\lim
}\limits_{\beta  \to 0} \left( {\widetilde K} \right)} \right) =
K_{Plane}= \overline {{E_S}} \frac{{4{{\left( {1 - {\nu _S}}
\right)}^2}}}{{3 - 4{\nu _S}}}\frac{\pi }{\lambda },
\end{equation}
in agreement with \cite{Audoly08a}.

\section{Terms of the linear stability analysis}\label{Appendix Linear analysis of stability}

The terms $a_i$, $\widetilde {{a_i}}$, $b_i$ and $\widetilde {{b_i}}$
that appear in Eq.~(\ref{eqPm}) are
\begin{subequations}
\begin{align}
{a_1} &=  - 4{\mkern 1mu} \xi {\mkern 1mu} h\left( {\left( {{m^4} +
\frac{3}{2} - \frac{{11}}{4}{\mkern 1mu} {m^2}} \right){h^2} - 9 -
3{\mkern 1mu} {m^2}} \right)\frac{\gamma }{{h + {{{h^3}}
\mathord{\left/
 {\vphantom {{{h^3}} {12}}} \right.
 \kern-\nulldelimiterspace} {12}} + {\mkern 1mu} {k_0}}},\\
\widetilde {{a_1}} &= h\left( {{{\left( {{m^2} - 1} \right)}^2}{h^2}
+ 12} \right)\xi  + 6{\mkern 1mu} \left( {1 - {\nu _S}}
\right){S_A},\\
{b_1} &=  - 4{\mkern 1mu} \xi {\mkern 1mu} h\left(
{\frac{1}{4}{\mkern 1mu} {h^2}{m^3} + 12{\mkern 1mu} m}
\right)\frac{\gamma }{{h + {{{h^3}} \mathord{\left/
 {\vphantom {{{h^3}} {12}}} \right.
 \kern-\nulldelimiterspace} {12}} + {\mkern 1mu} {k_0}}},\\
\widetilde {{b_1}} &=  - 12{\mkern 1mu} \xi {\mkern 1mu} hm +
6{\mkern 1mu} \left( {1 - {\nu _S}} \right){S_B},\\
{a_2} &=  - \left( {{h^3}{m^2} + 48{\mkern 1mu} h + 12{k_0}\left( {1
- {\mkern 1mu} \chi } \right)} \right)m\xi {\mkern 1mu} \frac{\gamma
}{{h + {{{h^3}} \mathord{\left/
 {\vphantom {{{h^3}} {12}}} \right.
 \kern-\nulldelimiterspace} {12}} + {\mkern 1mu} {k_0}}} + 12\gamma {\mkern 1mu} \left( {1 - \chi }
 \right)m\xi,\\
\widetilde {{a_2}} &=  - 12hm\xi  + 6{\mkern 1mu} \left( {1 - {\nu
_S}}
 \right){S_B},\\
{b_2} &= \left( {36{\mkern 1mu} h{m^2} + {h^3} + 12{\mkern 1mu} h +
12{\mkern 1mu} {k_0}{\mkern 1mu} \left( {1 - \chi } \right)}
\right)\xi \frac{\gamma }{{h + {{{h^3}} \mathord{\left/
 {\vphantom {{{h^3}} {12}}} \right.
 \kern-\nulldelimiterspace} {12}} + {\mkern 1mu} {k_0}}} - 12\gamma {\mkern 1mu} {\mkern 1mu} \left( {1 - \chi }
 \right)\xi,\\
\widetilde {{b_2}} &= 12h{m^2}\xi  + 6{\mkern 1mu} \left( {1 - {\nu
_S}} \right){T_B}
\end{align}
\end{subequations}
where $\gamma$, $k_0$, $S_A$, $S_B$ and $T_B$ were given in
\ref{Appendix Response of the substrate}.

\section{Influence of the cavity size}\label{Appendix Cavity size}

\begin{figure}[h!]
    \begin{center}
\includegraphics[width=\columnwidth]{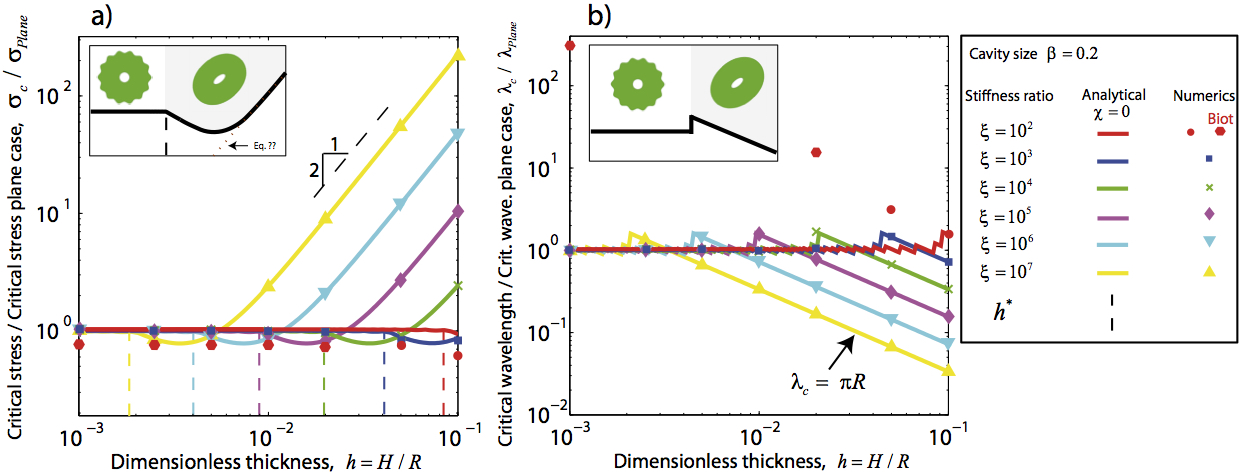}
\end{center}
\caption{(a) Critical stress $\sigma_c$ normalized by its planar
substrate counterpart $\sigma_{Plane}$, as a function of the
dimensionless ring thickness $h=H/R$. Analytical prediction is given
by Eq.~(\ref{Eq_Comparison_Plane_Sigma}). (b) Critical wavelength
$\lambda_c$ normalized by its planar substrate counterpart
$\lambda_{Plane}$, as a function of $h$. Analytical prediction is
given by Eq.~(\ref{Eq_Comparison_Plane_Lambda}). Insets are sketches
of the wrinkling and global buckling modes. The cavity size is
$\beta=0.2$ and the Poisson's ratio of the substrate is $\nu_S=0.5$.
The legend is common to both plots.}\label{Appendix_beta_02}
\end{figure}

\begin{figure}[h!]
    \begin{center}
\includegraphics[width=1.\columnwidth]{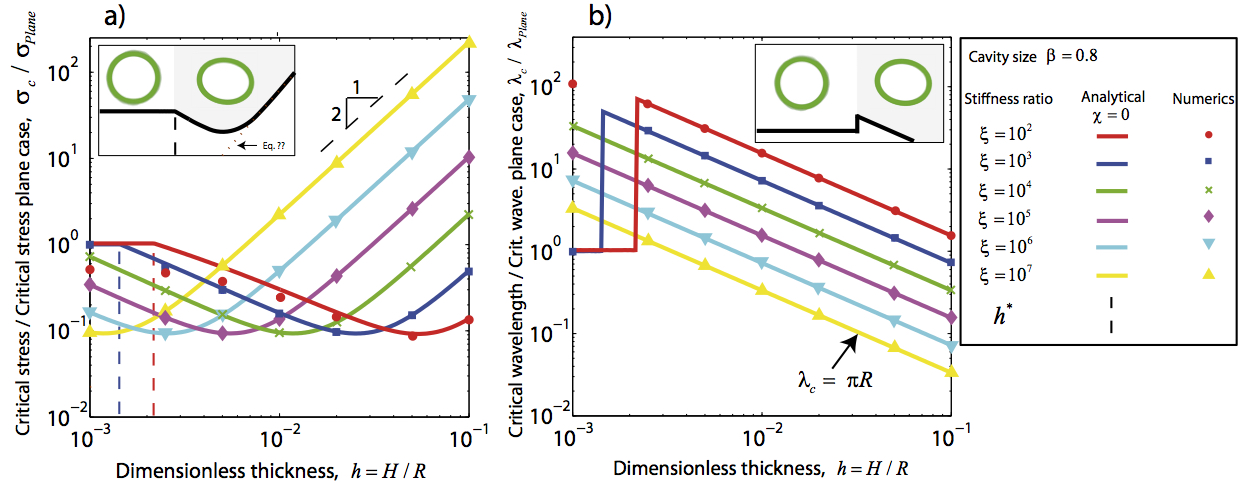}
\end{center}
\caption{(a) Critical stress $\sigma_c$ normalized by its plane
substrate counterpart $\sigma_{Plane}$, as a function of the
dimensionless ring thickness $h=H/R$. Analytical prediction is given
by Eq.~(\ref{Eq_Comparison_Plane_Sigma}). (b) Critical wavelength
$\lambda_c$ normalized by its plane substrate counterpart
$\lambda_{Plane}$, as a function of $h$. Analytical prediction is
given by Eq.~(\ref{Eq_Comparison_Plane_Lambda}). Insets are sketches
of the wrinkling and global buckling modes. The cavity size is
$\beta=0.8$ and the Poisson's ratio of the substrate is $\nu_S=0.5$.
The legend is common to both plots.}\label{Appendix_beta_08}
\end{figure}

In Figs.~\ref{Appendix_beta_02} and \ref{Appendix_beta_08}, we plot the $h$ dependence of $\sigma_c/\sigma_{Plane}$ and
$\lambda_c/\lambda_{Plane}$, for cavity sizes
$\beta=0.2$ and $\beta=0.8$, respectively. These plots are qualitatively similar to those obtained for $\beta=0.5$ in Figs. \ref{ComparisonPlaneStress} and
\ref{ComparisonPlaneWave}, and discussed in \S \ref{Section_Comparison_to_wrinkling_on_a_plane} of the main text.

\newpage

\bibliographystyle{apsrev4-1}
\bibliography{Biblio}

\end{document}